\documentclass[fleqn,usenatbib]{mnras}

\usepackage{newtxtext,newtxmath}

\usepackage{array}
\usepackage{subcaption}
\usepackage{amsmath,amssymb}
\usepackage{booktabs}
\usepackage{comment}
\usepackage{graphicx}

\usepackage{makecell}
\usepackage{xcolor}
\usepackage[T1]{fontenc}
\usepackage{hyperref}

\DeclareRobustCommand{\VAN}[3]{#2}
\let\VANthebibliography\thebibliography
\def\thebibliography{\DeclareRobustCommand{\VAN}[3]{##3}\VANthebibliography}

\newcommand{\orcid}[1]{\href{https://orcid.org/#1}{\textsuperscript{\includegraphics[width=8pt]{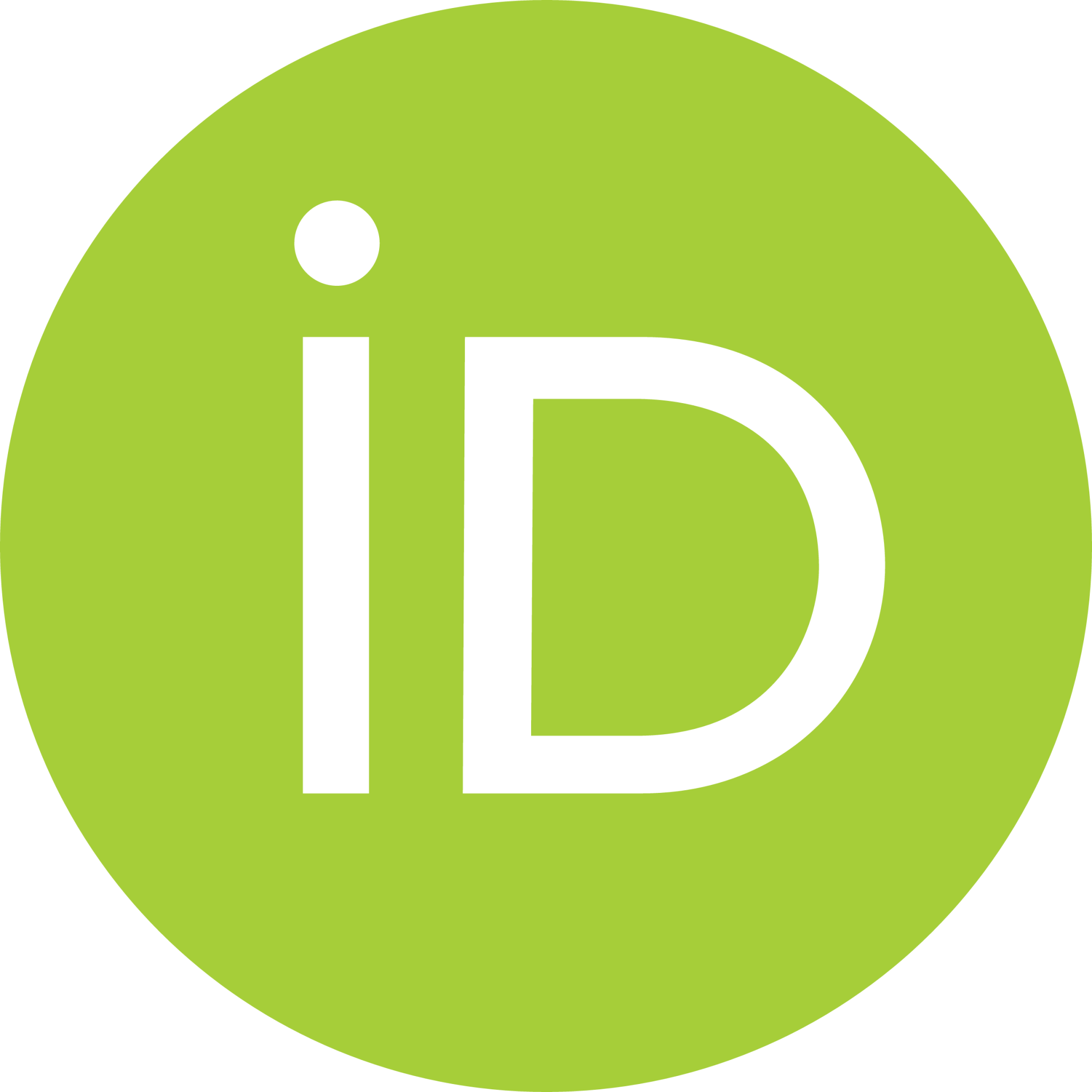}}}}

\usepackage{graphicx}	% Including figure files
\usepackage{amsmath}	% Advanced maths commands
\defcitealias{2025MNRAS.537.3245B}{B25}

\title[Equivalent width evolution]{A combined PRIMER+MINERVA study of the evolution 
of $\mathbf{[O\mbox{\sc{iii}}]+H\beta}$ equivalent width from $\mathbf{z=3}$ to $\mathbf{z=8.5}$}

\author[I. J. B. Holst]{Isaac J. B. Holst,$^{1}$\orcid{0009-0006-3003-9688} 
Ross J. McLure,$^{1}$\orcid{0009-0005-9742-2318} 
Ryan Begley,$^{2,1}$\orcid{0000-0003-0629-8074}
Derek J. McLeod,$^{1}$\orcid{0000-0003-4368-3326} 
James S. Dunlop,$^{1}$\orcid{0000-0002-1404-5950} 
\newauthor Patricia V. Fort,$^{1}$\orcid{0009-0008-1270-6218}
Stella Antonogiannaki,$^{1}$\orcid{0009-0006-2644-6424}
Cecilia Bondestam,$^{1}$\orcid{https://orcid.org/0009-0008-3775-5112}
Adam C. Carnall,$^{1}$\orcid{0000-0002-1482-5818}
Fergus Cullen,$^{1}$\orcid{0000-0002-3736-476X}
\newauthor Richard Ellis,$^{3}$
Norman A. Grogin,$^{4}$\orcid{0000-0001-9440-8872}
Anton M. Koekemoer,$^{4}$\orcid{0000-0002-6610-2048}
Ho-Hin Leung,$^{1}$\orcid{0000-0003-0486-5178}
\newauthor Feng-Yuan Frey Liu,$^{1}$\orcid{0000-0003-1386-3676}
Thomas M. Stanton,$^{1}$\orcid{0000-0002-0827-9769}
Struan D. Stevenson,$^{1}$\orcid{0000-0001-5642-752X}
Elizabeth Taylor$^{1}$\orcid{0000-0001-8728-2984}\\
$^{1}$Institute for Astronomy, University of Edinburgh, Royal Observatory, Edinburgh EH9 3HJ, UK\\
$^{2}$Armagh Observatory and Planetarium, College Hill, Armagh, BT61 9DG, N. Ireland, UK\\
$^{3}$Department of Physics and Astronomy, University College London, Gower Street, London WC1E 6BT, UK\\
$^{4}$Space Telescope Science Institute, 3700 San Martin Drive, Baltimore, MD 21218, USA\\
}

\date{Accepted XXX. Received YYY; in original form ZZZ}

\pubyear{\the\year{}}

\begin{document}
\label{firstpage}
\pagerange{\pageref{firstpage}--\pageref{lastpage}}
\maketitle

\begin{abstract}
    We present an analysis of the evolution of $\mbox{[O\,\sc{iii}]}+\mathrm{H}\beta$ equivalent width ($W_{\lambda}$) over the redshift range $z=3$ to $z=8.5$, based on combining the {\it JWST} NIRCam PRIMER and MINERVA imaging in the UDS survey field. We find that the combined medium- and broad-band data set reduces the photometric redshift outlier rate to $\eta = 4.7\%$ at $m_{\rm F444W}\leq28$, a $42\%$ reduction compared with PRIMER imaging alone.
    We also demonstrate that the combined data set facilitates the accurate recovery of $\mbox{[O\,\sc{iii}]}+\mathrm{H}\beta$ equivalent widths in redshift bins centred on $z\sim3.2,\, 4.0,\, 5.0,\, 7.2,\, 8.3$; corresponding to the F210M, F250M, F300M, F410M and F460M filters. 
    Based on a final sample of $N=1964$ galaxies with $M_{\rm UV}\leq-19.25$, we first provide new constraints on the bright end of the $z=7-8$ $\mbox{[O\,\sc{iii}]}+\mathrm{H}\beta$ luminosity function ($\log_{10}\phi^{*} = -4.3_{-0.5}^{+0.2}$, $\log_{10}L^{*} = -43.4_{-0.2}^{+0.3}$), before proceeding to investigate the redshift evolution of $\mbox{[O\,\sc{iii}]}+\mathrm{H}\beta$ equivalent width. Using SED modelling based on non-parametric continuity star-formation histories, we find that the median $\mbox{[O\,\sc{iii}]}+\mathrm{H}\beta$ equivalent width monotonically increases with redshift, rising from $\mathrm{log}_{10}(W_{\lambda})=2.54^{+0.01}_{-0.01}$ at $z\simeq 3$ to $\mathrm{log}_{10}(W_{\lambda})=3.05^{+0.03}_{-0.03}$ at $z\simeq 8$. In tandem, the distribution of $\mathrm{log}_{10}(W_{\lambda})$ is observed to become increasingly asymmetric with redshift, producing a fraction of extreme emission-line galaxies (EELGs, $W_{\lambda}>1000\,$\AA) that reaches $f_{\rm EELG}=0.50\pm{0.10}$ by $z=8.3$. Our SED modelling results indicate that the physical driver of this evolution is increasingly bursty star-formation, with EELGs the result of particularly high specific star-formation rates over the 10 Myr prior to observation.
\end{abstract}


\color{black}
\section{Introduction}
The photometric and spectroscopic capabilities of \textit{JWST} have established that galaxies at high redshift are synonymous with extremely strong nebular emission \citep[e.g.][]{Katz202521}. 
Given the comparative scarcity of extreme emission-line galaxies (EELGs) in the local Universe \citep[e.g.][]{amorin2015, 2021ApJ...922..170B}, understanding the evolution in nebular emission will provide direct insight into the star-formation rates, ages, metallicities and general interstellar medium (ISM) conditions of the high-redshift galaxy population \citep[e.g.][]{stark2016}. Moreover, nebular emission can also provide crucial information on the production, and subsequent escape, of ionizing photons from galaxies within the epoch of reionization.

The strength of rest-frame optical emission lines such as $\mbox{[O\,\sc{iii}]}\,\lambda\lambda4959, 5007$\AA\, reflects the production rate of massive O- and B-type stars on a timescale of $\simeq10$ Myr. This is in contrast to the optical continuum, which will typically reflect the contribution of the lower-mass, longer-lived stellar population. Consequently, the high equivalent widths first observed at $z\gtrsim6$ using a combination of \textit{Hubble Space Telescope} (HST) and \textit{Spitzer} photometry \citep[e.g.][]{Labbe_2013, Smit_2014,Roberts-Borsani_2016}, suggest that many high-redshift galaxies have recently experienced a rapid increase in their specific star-formation rate (sSFR).

This naturally leads to the conclusion that star formation in high-redshift galaxies is increasingly bursty/stochastic \citep{10.1093/mnras/stz2236, 10.1093/mnras/stz2286}, due to a combination of low stellar masses, high gas fractions and short dynamical timescales \citep[e.g.][]{10.1093/mnras/stx2595, 2025MNRAS.544.1732M}. Indeed, bursty star formation has been widely invoked to explain the high number of UV-bright galaxies seen at $z\gtrsim9$ \citep[e.g.][]{Castellano_2022, Naidu_2022, 10.1093/mnras/stac3472, 2023MNRAS.525.3254S, 2023MNRAS.521..497M, 2024ApJ...975..192G, 2024MNRAS.527.5004M} although, strictly, it appears that accounting for increasingly youthful stellar populations at earlier times is a sufficient explanation \citep[][]{2025MNRAS.539.2409D}.

In the bursty star-formation scenario, the nature of high-redshift galaxies that display low equivalent widths is potentially as interesting as that of extreme line emitters. Most of these galaxies are likely to be in a phase of reduced star-formation activity following a rapid burst, i.e. so-called mini-quenched galaxies \citep[e.g.][]{2024Natur.629...53L, 10.1093/mnras/staf213}. However, the low equivalent widths displayed by at least some of these galaxies could indicate ultra-low metallicities and/or a high escape fraction of Lyman continuum photons \citep[][]{Endsley_2025}.

Prior to the advent of \textit{JWST}, our ability to spectroscopically study rest-frame optical nebular emission was confined to $z\leq3.5$ by the available atmospheric windows in the near-IR \citep[e.g.][]{Kriek_2015, Holden_2016, Reddy_2018}. However, this situation was revolutionized by the availability of NIRSpec, which quickly made the detailed study of strong line-emitting galaxies at high redshift relatively straightforward \citep[e.g.][]{carnall23,10.1093/mnras/stad1408,borsani24}. 

At intermediate redshifts ($z\sim2$), the equivalent width ($W_{\lambda}$) of $\mbox{[O\,\sc{iii}]}$ is known to be highly correlated with ionizing photon production efficiency ($\xi_{\rm ion}$; \citealt{10.1093/mnras/stz2236}). Recent NIRSpec studies have found a similar $W_{\lambda}-\xi_{\rm ion}$ relation at $z\geq 6$, together with a strong correlation between [O{\sc iii}] $W_{\lambda}$ and $\beta$, the spectral slope of the UV continuum \citep[e.g.][]{begley26, papovich26}. Both correlations are consistent with the expectation that the nebular emission of high-redshift galaxies is dominated by a young, low-metallicity stellar population, with low dust attenuation.

Whilst the rest-frame optical nebular emission of individual high-redshift objects is best studied via spectroscopy, deep NIRCam imaging from legacy surveys such as PRIMER \citep{dunlopPRIMERPublicRelease2021}, CEERS \citep{2025ApJ...983L...4F} and JADES \citep{2026ApJS..283....6E} has provided an opportunity to study the evolution of the equivalent width of $\mbox{[O\,\sc{iii}]}+\mathrm{H}\beta$ for large, statistically complete samples. Recent photometric studies have demonstrated that extreme emission-line galaxies (EELGs: sources with [O{\sc iii}]+$\mathrm{H}\beta$ $W_{\lambda}\geq1000$\,\AA) were much more common in the early Universe, and that the typical $\mbox{[O\,\sc{iii}]}+\mathrm{H}\beta$ $W_{\lambda}$ increases from $\simeq 100$\,\AA\, at $z\simeq 2$ \citep{10.1093/mnras/stac1109} to $\simeq 700$\,\AA\, at $z\simeq 7.5$ \citep[e.g.][]{2025MNRAS.537.3245B,Endsley_2025}.

Until recently, photometric studies of $\mbox{[O\,\sc{iii}]}+\mathrm{H}\beta$ $W_{\lambda}$ have been somewhat limited by the lack of suitably deep, multi-band, \textit{JWST} medium-band imaging  over large areas. In this context, deep medium-band imaging in multiple filters is crucial because, compared to broad-band imaging alone, it has superior emission-line sensitivity and a greater ability to constrain
the underlying continuum level. The Medium band Imaging with NIRCam to Explore ReVolutionary Astrophysics \citep[MINERVA,][]{2025arXiv250719706M} survey aims to address this issue by observing areas of sky with existing NIRCam broad-band imaging in eight NIRCam medium-bands.
The large number of medium-band filters, when combined with previous broad-band data, opens up the opportunity for accurate and consistent nebular emission-line studies over a large range in redshift. Crucially, combining the MINERVA data with existing \textit{HST} and \textit{JWST} imaging in the PRIMER UDS and PRIMER COSMOS fields, allows photometric catalogues to be constructed featuring 19 filters, covering the wavelength range $0.4-5.0\,\mu$m.

Motivated by the completion of the MINERVA medium-band imaging in the PRIMER UDS field, we have exploited the combined PRIMER+MINERVA data set to consistently study the evolution of the $\mbox{[O\,\sc{iii}]}+\mathrm{H}\beta$ equivalent width distribution and the fraction of EELGs, within a $M_{\rm UV}-$complete galaxy sample spanning the redshift range $3<z<8.5$. As part of this process, we have also investigated the improvement in photometric redshift accuracy provided by the MINERVA medium-band imaging, and provide new constraints on the bright end of the $\mbox{[O\,\sc{iii}]}+\mathrm{H}\beta$ luminosity function at $z\simeq 7$. Based on the results of SED fitting using non-parametric star-formation histories, we also investigate the physical properties of emission-line galaxies as a function of redshift and $\mbox{[O\,\sc{iii}]}+\mathrm{H}\beta$ equivalent width.

The paper is structured as follows. In Section \ref{sec:data} we describe the production of the combined photometric catalogue, summarising the aperture photometry, point spread function (PSF) homogenization, dust corrections, and depth measurements. In Section \ref{sec:analysis} we describe the data analysis, including details of the photometric-redshift fitting, sample selection, the extraction of accurate equivalent width information from the photometry, and an estimate of the $z\simeq 7$ $\mbox{[O\,\sc{iii}]}+\mathrm{H}\beta$ luminosity function. We present our $\mbox{[O\,\sc{iii}]}+\mathrm{H}\beta$ emission-line results in Section \ref{sec:results} and discuss their implications in Section \ref{sec:disc}. Finally, we summarise our conclusions in Section \ref{sec:conclusions}. 

Throughout this paper we use the following cosmological parameters: $H_{0} = 70\,\rm{km}\,s^{-1}\,Mpc^{-1}$, $\Omega_{\rm{m}} = 0.3$, $\Omega_{\Lambda} = 0.7$, along with a \citet{10.1046/j.1365-8711.2001.04022.x} initial mass function (IMF). All magnitudes are presented in the AB system \citep{1983ApJ...266..713O}.

\section{Data}
\label{sec:data}
\subsection{Survey Imaging}
This work combines the \textit{JWST} PRIMER and MINERVA data sets with the existing {\it HST} ACS imaging \citep[Cosmic Assembly Near-infrared Deep Extragalactic Legacy Survey:][]{Grogin_2011, Koekemoer_2011} in the UDS survey field, to produce a joint photometry catalogue featuring a total of 19 wide- and medium-band filters. The PRIMER survey (GO-1837, PI: Dunlop, \citealt{dunlopPRIMERPublicRelease2021}; Magee et al., in preparation) was a {\it JWST} Cycle-1 Treasury Program providing deep photometric imaging observations in the CANDELS UDS and COSMOS fields. The PRIMER observations consist of imaging in eight NIRCam filters (F090W, F115W, F150W, F200W, F277W, F356W, F410M, F444W), with parallel imaging in two MIRI filters (F770W and F1800W). The MINERVA survey (GO-7814, PI: Muzzin, \citealt{2025arXiv250719706M}) is an on-going Cycle-4 program obtaining NIRCam medium-band imaging over an area of $\simeq 542$ $\mathrm{arcmin}^{2}$ in the UDS, COSMOS, EGS and GOODS-N survey fields. 

\begin{table}
    \caption{The $5\sigma$ global depths in all 19 filters. The depths are calculated in a 0.3$^{\prime\prime}-$diameter circular aperture and then corrected to total assuming a point-source morphology. The left-hand side lists the {\it HST} $^\dagger$ACS and PRIMER NIRCam filters, with the right-hand side listing the MINERVA NIRCam filters.}
    \centering
    \begin{tabular}{lccc}
    \hline
    Filter & $5\sigma_{\rm{global}}$ & Filter & $5\sigma_{\rm{global}}$ \\
    \hline
    F435W$\dagger$ & 27.39 & F140M & 27.10\\
    F606W$\dagger$ & 28.08 & F162M & 27.16\\
    F814W$\dagger$ & 28.09 & F182M & 27.43\\
    F090W & 27.71 & F210M & 27.25\\
    F115W & 27.75 & F250M & 27.33\\
    F150W & 27.95 & F300M & 27.69\\ 
    F200W & 28.12 & F360M & 27.69\\
    F277W & 28.38 & F460M & 26.71\\
    F356W & 28.37 & & \\
    F410M & 27.72 & & \\
    F444W & 27.83 & & \\
    \hline
    \end{tabular}
    
    \label{tab:glob_depths}
\end{table}

The first MINERVA field to be completed was the PRIMER UDS field, for which all observations were successfully taken by January 2026. These observations consist of NIRCam imaging in eight medium-band filters (F140M, F162M, F182M, F210M, F250M, F300M, F360M, F460M), together with MIRI parallel imaging in F1280W and F1500W. Both the PRIMER and MINERVA data sets were reduced with \textsc{pencil}: a bespoke version of the \textit{JWST} pipeline v1.17.1 (Magee et al., in preparation).

The area of the UDS field covered by the combined 16 NIRCam-filter imaging of PRIMER and MINERVA is $\sim$234 $\mathrm{arcmin}^{2}$, while an area of $\simeq 155\,\mathrm{arcmin}^{2}$ is also covered by all three {\it HST} ACS filters (F435W, F606W \& F814W). Below, we briefly detail the production of the combined photometry catalogue, with full details describing the production of combined PRIMER+MINERVA catalogues over both the PRIMER UDS and PRIMER  COSMOS fields deferred to a future paper (Holst et al., in preparation).

\begin{figure*}
    \centering
    \includegraphics[width=\textwidth]{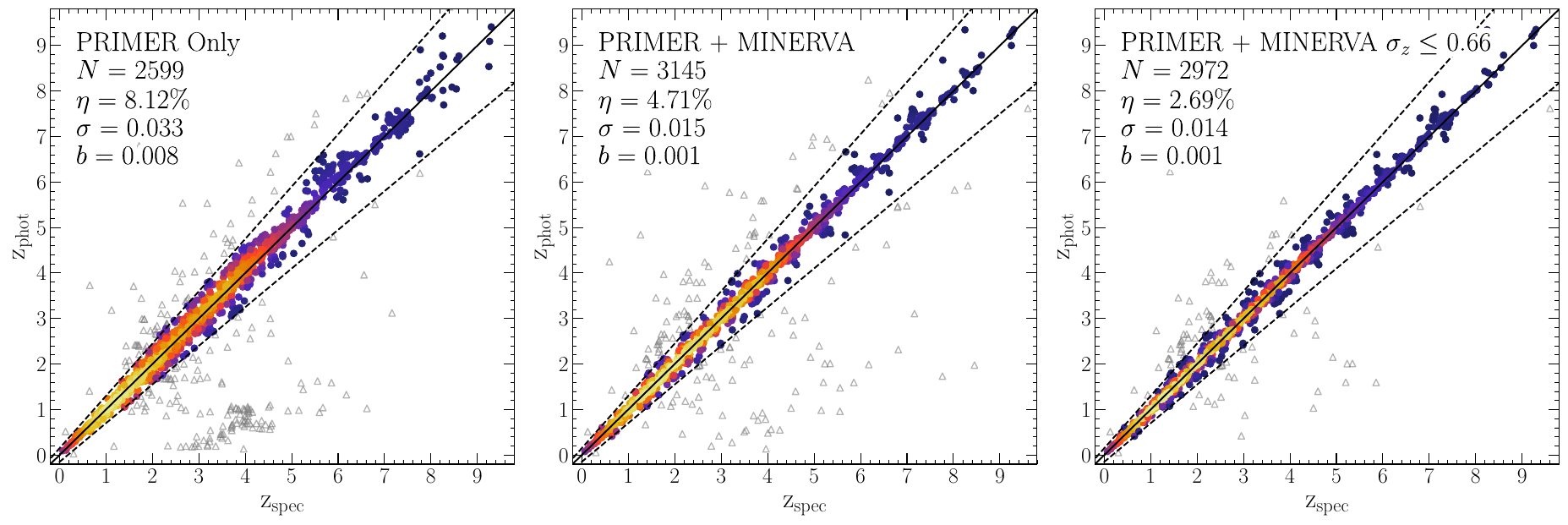}
    \captionsetup{skip=0.001pt}
    \caption{A comparison of photometric vs spectroscopic redshift for the PRIMER photometry (left), the PRIMER+MINERVA photometry (centre), and the PRIMER+MINERVA photometry including a $\sigma_{z}\leq0.66$ cut (designed to remove objects with photometric redshift distributions that have a high standard deviation, see Section \ref{sec:redshifts} for discussion). Black dotted lines show the outlier limits ($|\mathrm{d}z|\geq0.15$), outside of which an object is classified as a catastrophic outlier and plotted as a grey triangle. The non-outlier objects are coloured by the spatial density of points. We see a $42\%$ reduction in the outlier fraction ($\eta$) when the MINERVA photometry is added to the PRIMER photometry. We see a further 43\% reduction in the outlier rate when including the $\sigma_{z}\leq0.66$ cut.}
    \label{fig:photoz}
\end{figure*}

\subsection{PSF Homogenisation}
In order to obtain accurate photometry, each image was first PSF homogenised to match the angular resolution of the F444W imaging. For each filter, PSFs were initially constructed using stacks of bright, unsaturated and isolated stars. Transfer kernels were then calculated in Fourier space to match each filter's PSF to the F444W PSF, using varying window functions to reduce high-frequency artefacts such as fringing. Following convolution with the appropriate transfer kernel, the curve-of-growth in each filter matched that of the F444W imaging to within $\sim1\%$ at all radii.

\subsection{Photometry}

Photometry was measured using \textsc{SourceExtractor} \citep[][Version 2.28.0]{10.1046/j.1365-8711.1999.02978.x} in dual-image mode; using the unconvolved F356W image for object detection and the PSF-homogenised images to measure the photometry. The photometry was extracted in both 0.3$^{\prime\prime}$ and 0.5$^{\prime\prime}-$ diameter circular apertures, and corrected for galactic extinction using the \citet{Gordon_2023} extinction curve and $A_{V}~=~0.089$, as appropriate for the UDS field \footnote{https://dustmaps.readthedocs.io/en/latest/maps.html}. Robust, object-by-object photometry errors were calculated from the aperture-to-aperture rms of the nearest 200 blank sky apertures (using object masks based on the \textsc{SourceExtractor} segmentation maps). The median global depths for each of the filters in $0.3^{\prime\prime}-$diameter circular apertures (corrected to total) are presented in Table \ref{tab:glob_depths}.

\section{Analysis}
\label{sec:analysis}

\begin{table}
    \caption{The photometric redshift statistics for: (a) the total spectroscopic sample, (b) the ground-based spectroscopic sample and (c) the DJA spectroscopic sample. These results are shown with and without MINERVA photometry and with and without the redshift run standard deviation ($\sigma_{z}$) cut.}
    \centering
    \renewcommand{\arraystretch}{1.1}
    \begin{tabular}{ccccc}
    \multicolumn{5}{l}{\textbf{(a) Total}} \\
    \hline
    \textbf{Photometry} & $\mathbf{N}$ & $\boldsymbol{\eta}$ & $\boldsymbol{\sigma}$ & $\mathbf{b}$ \\
    \hline
    PRIMER & 2599 & 8.12\% & 0.033 & 0.008\\
    
    PRIMER+MINERVA & 3145 & 4.71\% & 0.015 & 0.001\\
    
    PRIMER+MINERVA $\sigma_{z}\leq0.66$ & 2972 & 2.69\% & 0.014 & 0.001\\
    \hline
    \multicolumn{5}{l}{}\\
    \multicolumn{5}{l}{\textbf{(b) Ground}} \\
    \hline
    \textbf{Photometry} & $\mathbf{N}$ & $\boldsymbol{\eta}$ & $\boldsymbol{\sigma}$ & $\mathbf{b}$ \\
    \hline
    PRIMER & 654 & 2.91\% & 0.023 & 0.004\\
    
    PRIMER+MINERVA & 654 & 0.92\% & 0.012 & -0.002\\
    
    PRIMER+MINERVA $\sigma_{z}\leq0.66$ & 650 & 0.92\% & 0.012 & -0.001\\
    \hline
    \multicolumn{5}{l}{}\\
    \multicolumn{5}{l}{\textbf{(c) DJA}} \\
    \hline
    \textbf{Photometry} & $\mathbf{N}$ & $\boldsymbol{\eta}$ & $\boldsymbol{\sigma}$ & $\mathbf{b}$ \\
    \hline
    PRIMER & 1945 & 9.87\% & 0.039 & 0.010\\
    
    PRIMER+MINERVA & 2491 & 5.70\% & 0.016 & 0.002\\
    
    PRIMER+MINERVA $\sigma_{z}\leq0.66$ & 2322 & 3.19\% & 0.015 & 0.002\\
    \hline
    
    \end{tabular}
    \label{tab:z_stats}
\end{table}

\begin{table*}
    \caption{The parameter ranges and priors used in the \textsc{bagpipes} SED fitting. $^{\dagger}$ This was changed to 0.00333 for the spectroscopic sample due to the higher reliability of these redshifts.}
    \centering
    \setlength{\tabcolsep}{23pt}
    \renewcommand{\arraystretch}{1.3}
    \begin{tabular}{cccc}
    \hline
    Parameter & Range & Prior & Hyper-parameters \\
    \hline
    Redshift & $\mu - 3\sigma \leq z_{\rm{phot}} \leq \mu + 3\sigma$ & Gaussian & $\mu = z_{\rm{phot}}, \; \sigma = 0.05^{\dagger}$\\
    Total stellar mass formed & $10^{6} \leq M_{*}/M_{\odot} \leq 10^{12}$ & $\mathrm{log}_{10}$ & \\
    Stellar Metallicity & $0.005 \leq Z_{*}/Z_{\odot} \leq 1$ & Uniform & \\
    $\rm{SFR\; Change}\, (i \rightarrow i+1)$ & $-10\leq\Delta_{i}\left(\mathrm{log}_{10}(\rm{SFR})\right) \leq 10$ & Student-t & $\rm{prior \; scale} = 1$, $\rm{prior \; d.f.} = 1.5$ \\
    V-band attenuation & $0\leq A_{V}\leq 3$ & Uniform & \\
    Nebular ionisation parameter & $-4 \leq \mathrm{log}(U) \leq -1$ & Uniform & \\
    \hline
    \end{tabular}
    \label{tab:bagpipes}
\end{table*}

\subsection{Photometric Redshifts}
\label{sec:redshifts}
Photometric redshifts were obtained using both \textsc{eazypy} \citep{Brammer_2008} and \textsc{LePhare} \citep[][using the \textsc{python} wrapper]{10.1046/j.1365-8711.1999.02978.x,ilbert2006photoz}. With \textsc{eazypy} we used four different template sets: two sets introduced in \citet{Larson_2023} and \citet{2024ApJ...964...71H} to model high-redshift galaxies, the "AGN\_BLUE\_SFHZ\_13" \footnote{https://github.com/gbrammer/eazy-photoz/tree/master/templates/sfhz} template set (which crucially contains the SED of a Little Red Dot from \citealt{killi2024lrd}) and an adaptation of the default \textsc{eazypy} template set with three additional high $W_{\lambda}$ emission-line templates \citep{2025MNRAS.537.3245B}. With \textsc{LePhare}, we adopted the "BC03\_CHAB\_MOD" \cite{10.1046/j.1365-8711.2003.06897.x} templates, along with the \textsc{PHYS} emission-line prescription.

The photometric redshift estimates were performed on both the 0.3$^{\prime\prime}$ and 0.5$^{\prime\prime}-$diameter aperture photometry. For the vast majority of sources ($87\%$ of the catalogue), the final photometric redshifts were based on the higher SNR 0.3$^{\prime\prime}-$diameter photometry. However, for bright objects ($m_{\mathrm{F356W}}<25.2$) we adopted the photometric redshifts derived from the 0.5$^{\prime\prime}-$diameter photometry. For this sub-set of sources, that are typically extended, testing revealed that the larger aperture delivered better colours and superior photometric-redshift accuracy.

For each object, we adopted the median of the five different redshift runs as the final photometric redshift. To evaluate their accuracy, we compared our photometric redshifts to a compilation of $N = 3145$ high-quality spectroscopic redshifts within the 234 arcmin$^{2}$ PRIMER+MINERVA footprint.
The majority of these (2491) were grade=3 spectroscopic redshifts from the Dawn JWST Archive \citep[DJA, ][]{valentino2025outflows, pollock2026aurora}. The remaining 654 spectroscopic redshifts were drawn from a variety of ground-based spectroscopic surveys including: VANDELS \citep{mclure18,garilli21}, UDSz \citep{bradshaw13,mclure13}, MOSDEF \citep{Kriek_2015} and \cite{galametz18}. The ground-based spectroscopic comparison sample is typical of the bright, blue samples of spectroscopic objects traditionally used to test the accuracy of photometric redshift fitting. In contrast, the sample of DJA spectra are fainter, redder and more typical of the galaxies featuring in our PRIMER+MINERVA photometry catalogue. 

To evaluate the performance of our photometric-redshift fitting, we measured: $\mathrm{d}z = (z_{\mathrm{phot}} - z_{\mathrm{spec}})/(1\; +\; z_{\mathrm{spec}})$, the catastrophic outlier fraction, $\eta$ (where $|\mathrm{d}z|\geq0.15$), the standard deviation, $\sigma = 1.483\times \rm{MAD}(\rm{d}z)$ and the bias, $b = \mathrm{Median}\left(z_{\mathrm{phot}} - z_{\mathrm{spec}}\right)$. The full set of statistics are reported in Table \ref{tab:z_stats} for those objects that have $\rm{SNR}\geq5$ in the F356W filter and benefit from the full suite of relevant filters. Based on the PRIMER NIRCam plus {\it HST} ACS photometry, our photometric redshifts have an overall accuracy of $\sigma=0.033$ and an outlier fraction of 8.1\%. However, when the medium-band data from MINERVA are included, the overall accuracy improves to $\sigma=0.015$ and the outlier fraction drops to 4.7\%. 
The inclusion of MINERVA's medium-bands therefore improves the photometric-redshift accuracy by more than a factor of two and reduces the catastrophic outlier rate by $42\%$. 

One benefit of calculating our photometric redshifts from five different runs is that we are able to easily identify objects where the five best-fitting redshifts have a large standard deviation ($\sigma_{z}$). As might be hoped, this provides a useful method for identifying objects that are more likely to be catastrophic outliers. In the final row of Table \ref{tab:z_stats}, we display the statistics for the PRIMER+MINERVA photometry with the additional criterion that $\sigma_z \leq 0.66$, resulting in an outlier rate of only $\eta = 2.69\%$, down to a magnitude limit of $m_{\rm F444W}\leq28$. This criterion is therefore capable of reducing the outlier fraction by a further 43\%, while excluding $\leq 7\%$ of the overall sample. Fig. \ref{fig:photoz} shows the comparison between photometric and spectroscopic redshifts for the three photometric samples presented in Table \ref{tab:z_stats}.

As expected, in all cases, the fidelity of the photometric redshifts is significantly better when tested against the bright, blue, ground-based spectroscopic sample, than the fainter, redder, DJA spectroscopic sample. That said, the improvement in photometric-redshift outlier rates for the more representative DJA spectroscopic sample is particularly significant. Using the photometric redshifts provided by the DJA (PRIMER data only) results in an outlier fraction of $12.9\%$, compared to our outlier fraction of $5.7\%$ (or $3.2\%$ including the $\sigma_{\rm z}$ cut). Moreover, isolating only objects from the RUBIES survey (PID-4233; PIs: de Graaff \& Brammer), we measure $\eta = 3\%$; compared to the $\eta=6\%$ obtained by \citet{degraaff2025rubies} using only PRIMER data.

\subsection{$\mbox{[O\,\sc{iii}]} + \rm{H}\beta$ Equivalent Widths}
\label{sec:ew_methods}
In addition to improving overall photometric redshift accuracy, the inclusion of the MINERVA medium-band imaging makes it possible to study nebular line emission over a wide range in redshift. In particular, the availability of F460M imaging can constrain the $\mbox{[O\,\sc{iii}]}+\mathrm{H}\beta$ equivalent width out to $8\leq z\leq8.5$, while the F410M PRIMER imaging provides an opportunity to benchmark our results against previous studies at $z\simeq7$ \citep[e.g.][]{10.1093/mnras/stae2353, 2025MNRAS.537.3245B}. 

To determine the $\mbox{[O\,\sc{iii}]} + \rm{H}\beta$ equivalent widths ($W_{\lambda}$) we modelled the 19-filter photometry data set with \textsc{bagpipes} \citep[Bayesian Analysis of Galaxies for Physical Inference and Parameter EStimation, ][]{10.1093/mnras/sty2169}, closely following the methodology described in \cite{2025MNRAS.537.3245B}. We tested both the Binary Population and Spectral Synthesis (BPASS) v2.2.1 \citep{Eldridge_Stanway_Xiao_McClelland_Taylor_Ng_Greis_Bray_2017, 10.1093/mnras/sty1353} and BC03 \citep[][updated 2016 version]{10.1046/j.1365-8711.2003.06897.x} stellar population synthesis (SPS) models, along with a variety of different star formation histories (SFHs): delayed-$\tau$ models (see e.g. \citealt{2019ApJ...873...44C} for parametric model analysis), constant, and an assortment of hyper-parametrisations of the non-parametric continuity SFH presented in \citet{Leja_2019}.

To fine tune our \textsc{bagpipes} setup we used \textsc{specutils} 
to measure the $\mbox{[O\,\sc{iii}]} + \rm{H}\beta$ equivalent widths for 115 galaxies drawn from our full sample that have publicly available NIRSpec PRISM spectra on the DJA. These sources have SNR $\geq5$ in the F356W filter and $M_{\rm{UV}}\leq-19.25$, to match the constraints we use in Section \ref{sec:selection} for photometric selection. We allow the redshift of the galaxies to vary with a Gaussian prior ($\sigma = 0.00333$). To ensure accurate spectroscopic measurements of the $\mbox{[O\,\sc{iii}]} + \rm{H}\beta$ equivalent widths, these spectra were chosen to have SNR $\geq5$ over the $\mbox{[O\,\sc{iii}]} + \rm{H}\beta$ complex, together with significant detections of the adjacent continuum: 
median SNR/pix $\geq1.5$ in the continuum fitting regions. 

The best recovery of the spectroscopic equivalent widths were delivered
by a non-parametric continuity SFH, using the BPASS SPS models and the {\sc bagpipes} default BPASS \textsc{cloudy} nebular emission models \citep{2013RMxAA..49..137F}. To make the \citet{Leja_2019} SFH allow larger variations in SFR, we increased the \textsc{$\Delta_{i}\left(\mathrm{log}_{10}(\rm{SFR})\right)$ prior scale} from 0.3 to 1, and decreased the \textsc{$\Delta_{i}\left(\mathrm{log}_{10}(\rm{SFR})\right)$ prior d.f.} from 2 to 1.5. These hyper-parameters dictate $\sigma$ and $\nu$, respectively, in the Student's t-distribution, with the chosen values similar to those used in previous studies \citep[see e.g.][]{2022ApJ...927..170T, 2025MNRAS.537.3245B}. We use age bins of $0-3, \, 3-10,\, 10-100, \, 100-250, \,\mathrm{and}\; 250-500\, \mathrm{Myr}$. The full set of \textsc{bagpipes} parameters used in our final analysis is presented in Table \ref{tab:bagpipes}. 

\begin{figure}
    \centering
    \includegraphics[width=\linewidth]{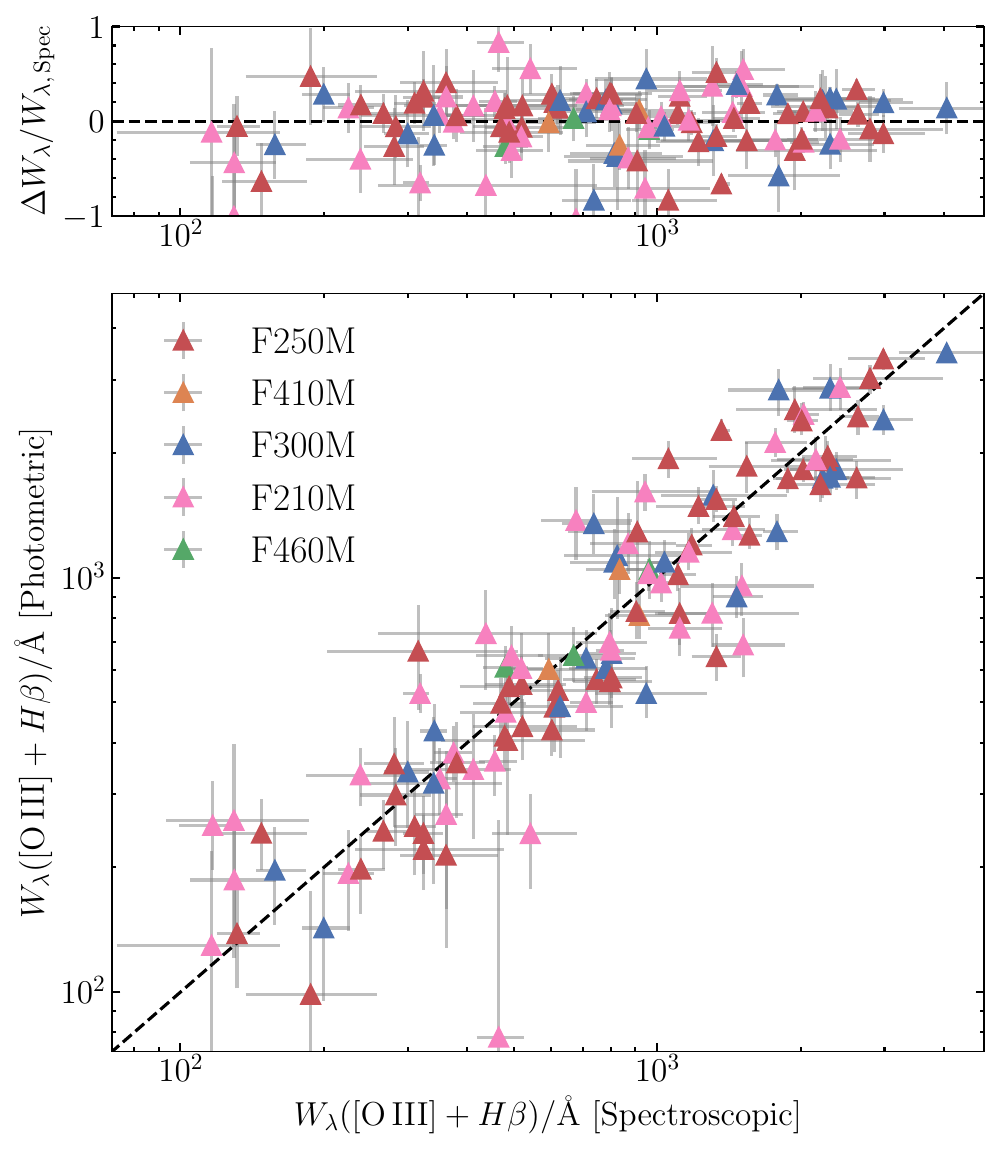}
    \caption{Photometric versus spectroscopic measurements of rest-frame $\mbox{[O\,\sc{iii}]} + \rm{H}\beta$ equivalent width for our sub-sample of $N=115$ galaxies with high-quality NIRSpec PRISM spectra (see Section \ref{sec:ew_methods}). The dashed line shows the 1:1 relation. The data points are coloured by the filter within which the emission lines were observed (F460M in green, F410M in orange, F300M in blue, F250M in red, and F210M in pink). The upper panel shows the relative difference in $W_{\lambda}$ against the spectroscopic $W_{\lambda}$. We find a relationship of $W_{\lambda, \, \rm{spec}} = (1.04\pm 0.04)W_{\lambda, \, \rm{phot}}$.\color{black}}
    \label{fig:ew_spec}
\end{figure}

For each object, we obtain the line and continuum fluxes directly from the \textsc{bagpipes} output parameter posteriors, generating a distribution of $N_{\rm{posterior}} = 1000$ equivalent widths. We take the $50^{\rm{th}}$ percentile of the resulting distribution to be our $W_{\lambda}$ measurement, and the $16^{\rm{th}}$ and $84^{\rm{th}}$ percentiles as our $1\sigma$ errors. 
Fig. \ref{fig:ew_spec} shows the comparison of our photometric and spectroscopic equivalent widths, coloured by the filter in which the emission lines were observed. The relationship between photometric and spectroscopic equivalent widths is close to one-to-one, with a scatter of $\sigma(W_{\lambda})=337$\,\AA. To account for the minor discrepancy between our photometric measurements and their spectroscopic counterparts, we fit the data with a linear relationship, finding $W_{\lambda, \, \rm{spec}} = (1.04\pm 0.04)W_{\lambda, \, \rm{phot}}$. All future $W_{\lambda}$ measurements are corrected using this relation.

\subsection{Sample Selection}
\label{sec:selection}
Our initial sample of galaxies comprised objects with $\rm{SNR}\geq5$ in the F356W filter and a photometric redshift in the ranges $2.95 \leq z \leq 3.4$, $3.83 \leq z \leq 4.2$, $4.65 \leq z \leq 5.3$, $6.7 \leq z \leq 7.6$,  or $8.0 \leq z \leq 8.5$, corresponding to the $\mbox{[O\,\sc{iii}]} + \rm{H}\beta$ complex being redshifted into the F210M, F250M, F300M, F410M and F460M filters. In each redshift bin, the sample was further restricted to those objects with $\rm{SNR}\geq3$ in the corresponding medium-band filter and a careful visual inspection was performed to remove any remaining artefacts. In this study we deliberately avoid using the F360M filter to measure $\mbox{[O\,\sc{iii}]} + \rm{H}\beta$ equivalent widths in the redshift range $5.8 \leq z \leq 6.65$. This is because
the F360M filter sits within the broad F356W filter, meaning that our parent photometry catalogue is biased towards high $\mbox{[O\,\sc{iii}]} + \rm{H}\beta$ equivalent width objects.

To exclude potential active galactic nuclei (AGN) we removed $N=13$ sources detected in X-rays within $0.5^{\mathrm\prime\prime}$ \citep[using][]{Kocevski_2018}, and $N=35$ sources in common with the Little Red Dot (LRD) samples of \citet{Kocevski_2023} and \citet{perger2025lrd}. At this stage, our galaxy sample consists of $N=5325$ objects, over the redshift range $2.95<z<8.50$. However, in order to mitigate the evolving $M_{\rm UV}$ limit, we restrict our final sample to those objects with $M_{\rm{UV}}\leq-19.25$, after conservatively estimating that our sample is complete to this limit across the full redshift range. The final sample, used for all subsequent analysis, consists of $N=1964$ objects, with $N=633$ in the $2.95 \leq z \leq 3.4$ bin, $N=584$ in the $3.83 \leq z \leq 4.2$ bin, $N=430$ in the $4.65 \leq z \leq 5.3$ bin, $N=268$ in the $6.7 \leq z \leq 7.6$ bin and $N=49$ in the $8.0 \leq z \leq 8.5$ bin.

\begin{figure}
    \centering
    \includegraphics[width=\linewidth]{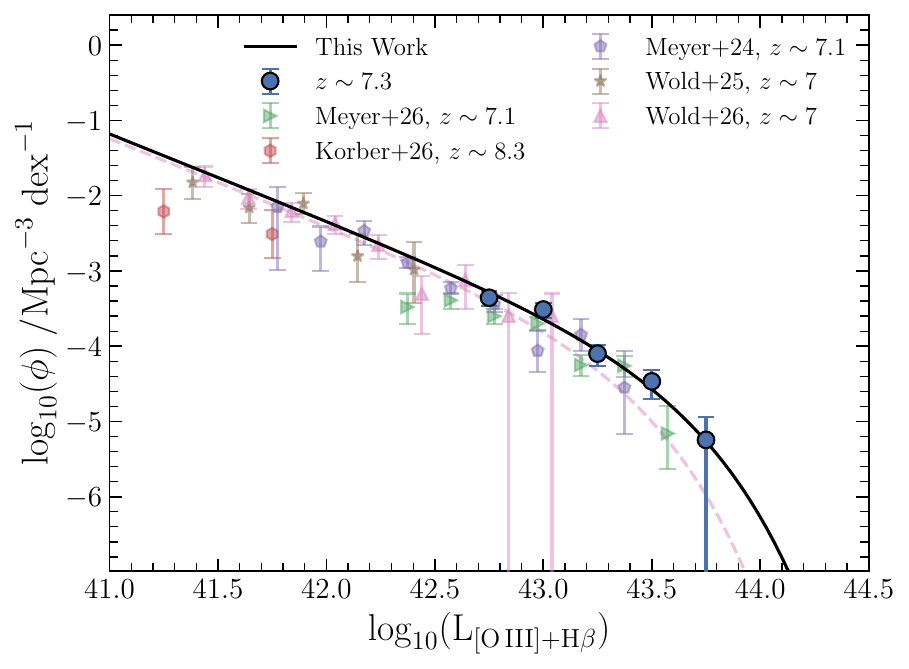}
    \caption{A new estimate of the bright-end of the $\mbox{[O\,\sc{iii}]} + \rm{H}\beta$ luminosity function at redshift $z\simeq 7.3$ (filled blue circles), based on combining our $8.0\leq z <8.5$ and $6.7\leq z <7.6$ sub-samples. We also show two LF estimates at $z\simeq 7.1$ from \citet{10.1093/mnras/stae2353} and \citet{2026A&A...712A..16M}, an LF estimate at $z\simeq 7$ by \citet{2025ApJ...980..200W} and the two highest luminosity data points from the $z\simeq8.3$ LF estimate of \citet{korber2026glimpse}. Where necessary we have corrected from $\mbox{[O\,\sc{iii}]}\lambda5007$ to $\mbox{[O\,\sc{iii}]} + \rm{H}\beta$ luminosities. The pink dashed line shows the Schechter function fit at $z\simeq 7$ from \citet{2026arXiv260605323W}. The solid black line shows a combined fit to our new bright-end data and the fainter \citet{2026arXiv260605323W} data (see Section \ref{sec:lf} for details).}
    \label{fig:oiii_lf}
\end{figure}

\begin{figure}
    \centering
    \includegraphics[width=\linewidth]{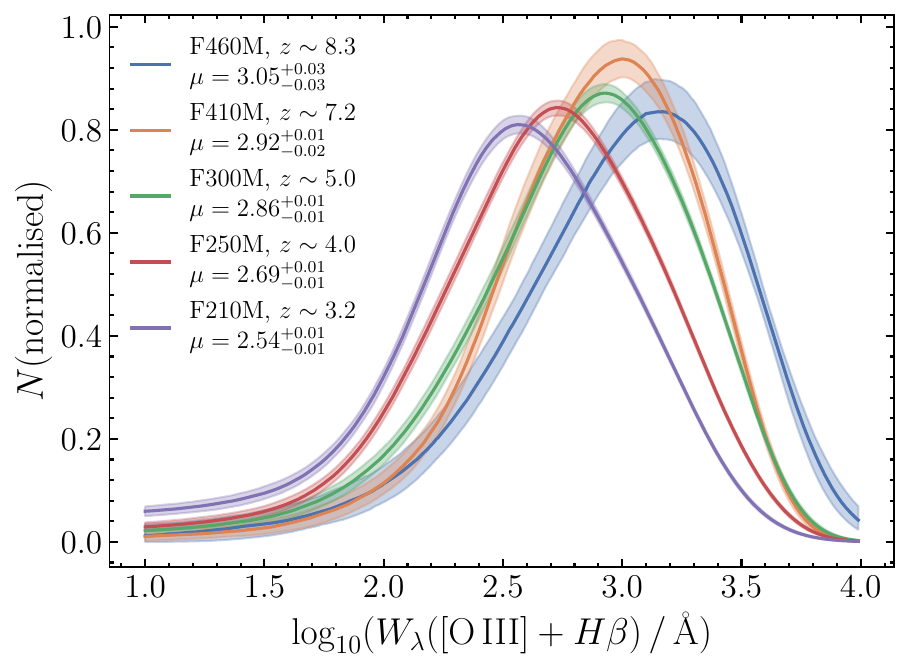}
    \caption{Kernel density estimates of the $\mbox{[O\,\sc{iii}]} + \rm{H}\beta$ equivalent width distributions for the F210M (purple, $2.95 \leq z \leq 3.4$), F250M (red, $3.83 \leq z \leq 4.2$), F300M (green, $4.65 \leq z \leq 5.3$), F410M (orange, $6.7 \leq z \leq 7.6$) and F460M (blue, $8.0 \leq z \leq 8.5$) sub-samples. The shaded regions surrounding the lines are the $\pm 1\sigma$ regions calculated by Monte Carlo sampling of the individual posterior distributions. The higher redshift samples show a deviation from the log-normal shape seen at lower redshifts \citep[e.g.][]{Lee_2007, Ly_2011, Stark_2013, Schenker_2014, 10.1093/mnras/stac1109, 2025MNRAS.537.3245B}.}
    \label{fig:ew_dist}
\end{figure}

\begin{figure}
    \centering
    \includegraphics[width=\linewidth]{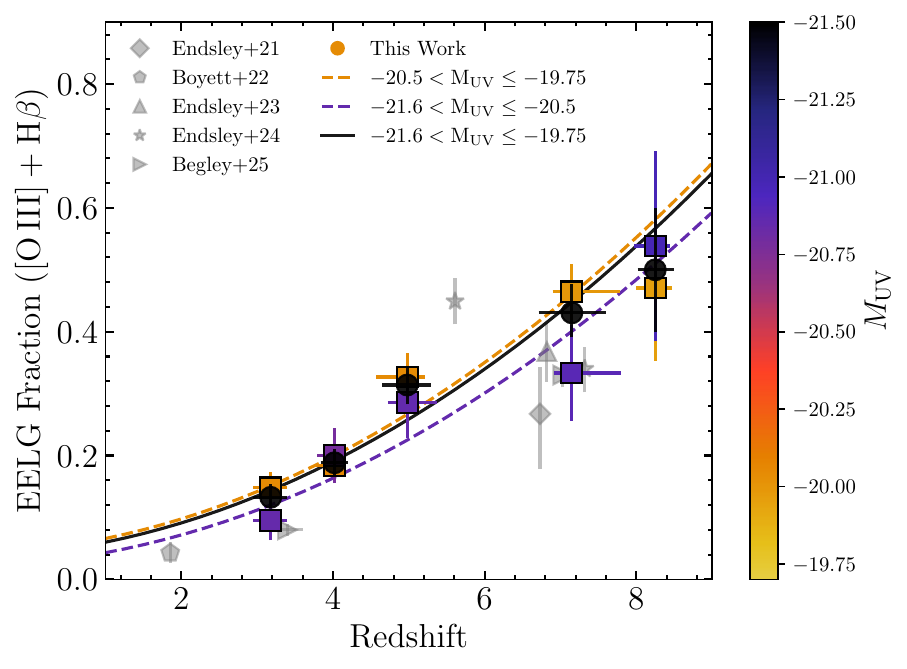}
    \caption{The fraction of extreme emission-line galaxies (EELGs) against redshift. The EELG fractions for the full $-21.60<M_{\mathrm{UV}}\leq -19.75$ sample are shown as filled black circles. The EELG fractions for bright ($-21.60<M_{\mathrm{UV}}\leq -20.5$) and faint ($-20.5<M_{\mathrm{UV}}\leq -19.75$)
    sub-samples are shown as filled squares, coloured by
    their median $M_{\rm{UV}}$. The solid and dashed lines show quadratic fits to our full sample, and the bright and faint $M_{\rm UV}$ sub-samples, respectively. Also shown are results from \citet{10.1093/mnras/staa3370}, \citet{10.1093/mnras/stac1109}, \citet{10.1093/mnras/stad1919}, \citet{10.1093/mnras/stae1857},  and \citet{2025MNRAS.537.3245B}. The literature results cover a similar range of UV luminosities to our data and have been corrected to a common EELG definition. The observed quadratic increase with redshift aligns with an increase in star-formation burstiness at high-redshift.}
    \label{fig:eelg}
\end{figure}

\subsection{$\mbox{[O\,\sc{iii}]} + \rm{H}\beta$ Luminosity Function}
\label{sec:lf}
As a final consistency check on our sample selection and SED-fitting methodology, we computed the $\mbox{[O\,\sc{iii}]} + \rm{H}\beta$ luminosity function (LF) at $z\simeq 7.3$ by combining our F410M and F460M sub-samples. Our LF estimate is based on the $1/V_{\rm{max}}$ method \citep{1968ApJ...151..393S}, with final errors calculated as the quadrature sum of bootstrap resampling and cosmic variance \citep{2008ApJ...676..767T} uncertainties.

In Fig. \ref{fig:oiii_lf} we show our new estimate of the bright-end of the LF at $z\simeq 7.3$, together with recent literature results at $z\simeq 7.1-8.3$ from \citet{10.1093/mnras/stae2353}, \citet{2026A&A...712A..16M}, \citet{2025ApJ...980..200W}, \citet{2026arXiv260605323W}, and \citet{korber2026glimpse}. The \citet{1976ApJ...203..297S} function fit to the $z\simeq7$ LFs by \citet{2026arXiv260605323W} is also shown as a pink dotted line. 
Where necessary, we have corrected $\mbox{[O\,\sc{iii}]}\lambda5007$ luminosities to $\mbox{[O\,\sc{iii}]} + \rm{H}\beta$ luminosities using a factor of 1.48 \citep{2025MNRAS.537.3245B}. Our new LF estimate is plotted at $\log_{10} L_{\mbox{[O\,\sc{iii}]} + \rm{H}\beta}\geq 42.6$, which corresponds to our conservative completeness limit, based on the $5\sigma$ limiting line flux from the \textsc{bagpipes} SED modelling and the turnover in the number counts at fainter luminosities.

Our luminosity function estimate agrees well with the $z = 7$ LF of \citet{2026arXiv260605323W} and also follows cleanly on from the $z = 7$ results of \citet{2025ApJ...980..200W}. Our bright-end constraints are also consistent with, albeit marginally higher than, the data points from \citet{10.1093/mnras/stae2353} and \citet{2026A&A...712A..16M} at $z\simeq 7.1$, and the Schechter function fit from \citet{2026arXiv260605323W}.

Due to our completeness limit, our new results provide no constraint on the faint-end slope of the LF. We therefore combine our data with that of \citet{2026arXiv260605323W} (disregarding their brightest two single-object bins) and fit the combined data set at $\log_{10} L_{\mbox{[O\,\sc{iii}]} + \rm{H}\beta}\geq 41.4$. Our best-fitting Schechter function is shown as the solid black line in Fig. \ref{fig:oiii_lf} and has the following parameters: $\log_{10}(\phi^{*}/\rm{Mpc^{-3}\,dex^{-1}}) = -4.3_{-0.5}^{+0.2}$, $\log_{10}(L^{*}/\rm{erg\, s^{-1}\,dex^{-1}}) = -43.4_{-0.2}^{+0.3}$, and $\alpha = -2.15^{+0.05}_{-0.07}$. 

It can be seen that our new Schechter function is 0.5 dex higher than the \citet{2026arXiv260605323W} fit at $\log_{10} L_{\mbox{[O\,\sc{iii}]} + \rm{H}\beta}\geq 43.75$. This could be a result of the different observation methods, with \citet{2026arXiv260605323W} using the slitless-spectroscopy of \citet{10.1093/mnras/stae2353} to constrain the bright end of their LF. It will be possible to confirm or refute this result with the larger area that will be provided by the final complete MINERVA + PRIMER/CEERS/JADES data set.

\section{Results}
\label{sec:results} 
\begin{figure*}
    \centering
    \captionsetup[subfigure]{labelformat=empty, skip = 0.001pt}
    \begin{subfigure}[t]{0.47\textwidth}
    \centering
    \includegraphics [width=\textwidth]{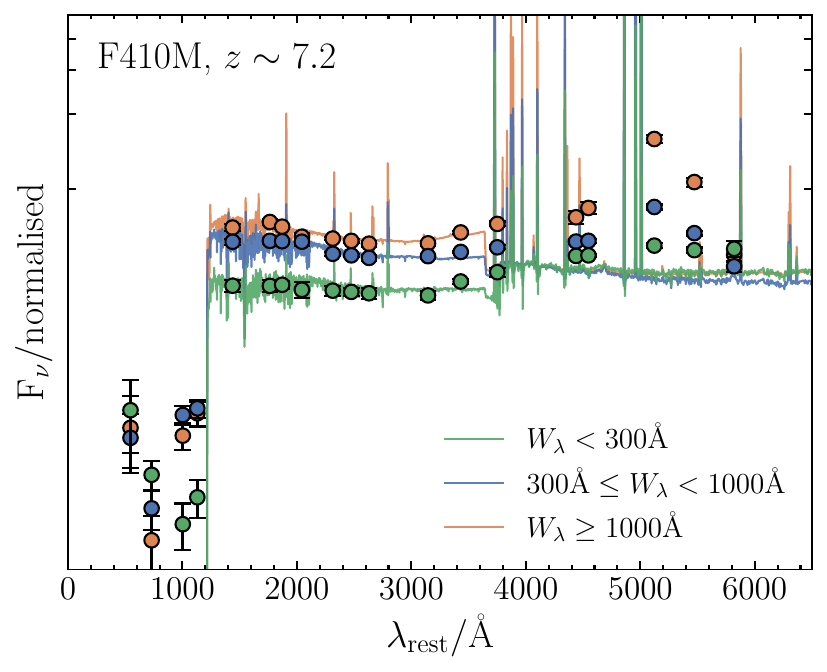}
    \caption{}
    \label{fig:f460_sed}
    \end{subfigure}
    \begin{subfigure}[t]{0.47\textwidth}
    \centering
    \includegraphics[width = \textwidth]{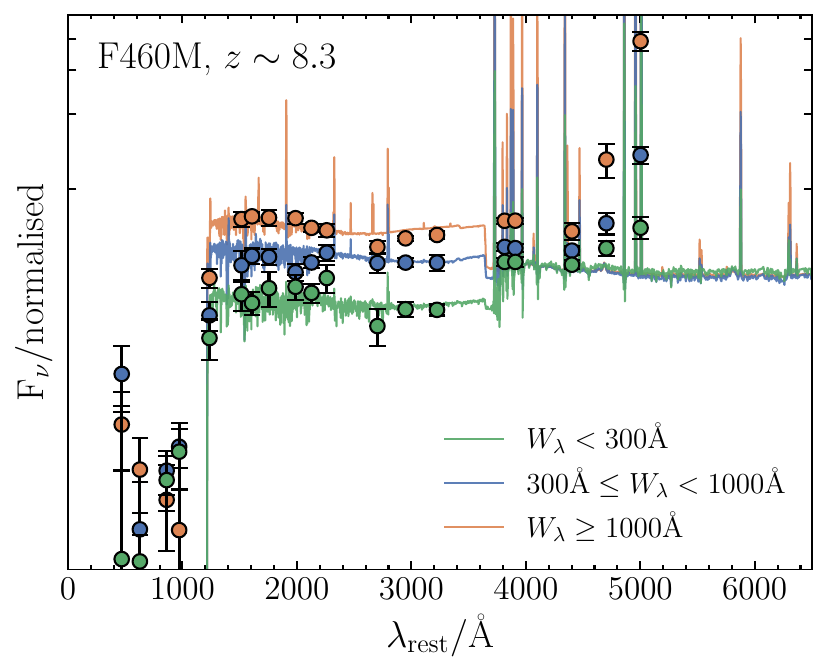}
    \caption{}
    \label{fig:f410_sed}
    \end{subfigure}
    \captionsetup{skip=0.001pt}
    \caption{Stacked and normalised SEDs for the F410M sample at $6.7\leq z\leq 7.6$ (left) and the F460M sample at $8.0\leq z\leq 8.5$ (right). Stacked SEDs for galaxies with $W_{\lambda}\geq1000\,$\AA\,are shown in orange, with stacked SEDs for galaxies with $300\leq W_{\lambda}<1000\,$\AA\, and $W_{\lambda}<300\,$\AA\, shown in blue and green, respectively. Stacked photometry for each sub-sample is overlaid as filled circles of the appropriate colour.}
    \label{fig:stack_sed}
\end{figure*}

In Section \ref{sec:ew_methods} we demonstrated that detailed SED fitting of the combined PRIMER+MINERVA data set provides accurate emission-line measurements for $z\simeq 3-8.5$ galaxies. In this section we investigate the evolving $\mbox{[O\,\sc{iii}]} + \rm{H}\beta$ equivalent width distributions and underlying physical properties of our galaxy samples at $z\simeq3.2$, $z\simeq4.0$, $z\simeq5.0$, $z\simeq7.2$ and $z\simeq8.3$.

As $\mbox{[O\,\sc{iii}]} + \rm{H}\beta$ equivalent width provides a direct probe of nebular emission and recent star-formation, tracing its evolution should offer insight into changes in the typical star-formation history of our sample as function of redshift. Moreover, by focusing on the prevalence and physical properties of EELGs, it should be possible to obtain a clearer picture of the importance of bursty star formation at high redshift.

\subsection{$W_{\lambda}$ Distributions}
\label{sec:EW_dists}
We present the distributions of photometric $\mbox{[O\,\sc{iii}]} + \rm{H}\beta$ equivalent widths for all redshift samples in Fig. \ref{fig:ew_dist}. For each object in the samples, a random $W_{\lambda}$ was chosen from its posterior distribution, and a Gaussian kernel density estimate (KDE) was taken of the random draws to better visualise the underlying distribution. This method was repeated 1000 times, creating a posterior distribution of normalised KDE counts at each value of $\mathrm{log}_{10}(W_{\lambda})$, from which the $16^{\rm th}, \; 50^{\rm th}, \;\rm{and}\; 84^{\rm th}$ percentiles were taken. The median values of $\mathrm{log}_{10}(W_{\lambda})$ for the F210M, F250M, F300M, F410M, and F460M samples are listed in the Fig. \ref{fig:ew_dist} legend. It is clear that the median value of $\mathrm{log}_{10}(W_{\lambda})$ is monotonically increasing with redshift. Moreover, it can also be seen that the $W_{\lambda}$ distributions become increasingly asymmetric at the highest redshifts, with longer tails towards low equivalent widths. We discuss the implications of these results, along with comparisons to previous literature studies in Section \ref{sec:EW_disc}

\begin{figure}
    \centering
    \captionsetup[subfigure]{labelformat=empty, skip = 0.001pt}
    \begin{subfigure}[t]{0.48\textwidth}
    \centering
    \includegraphics[width=\textwidth]{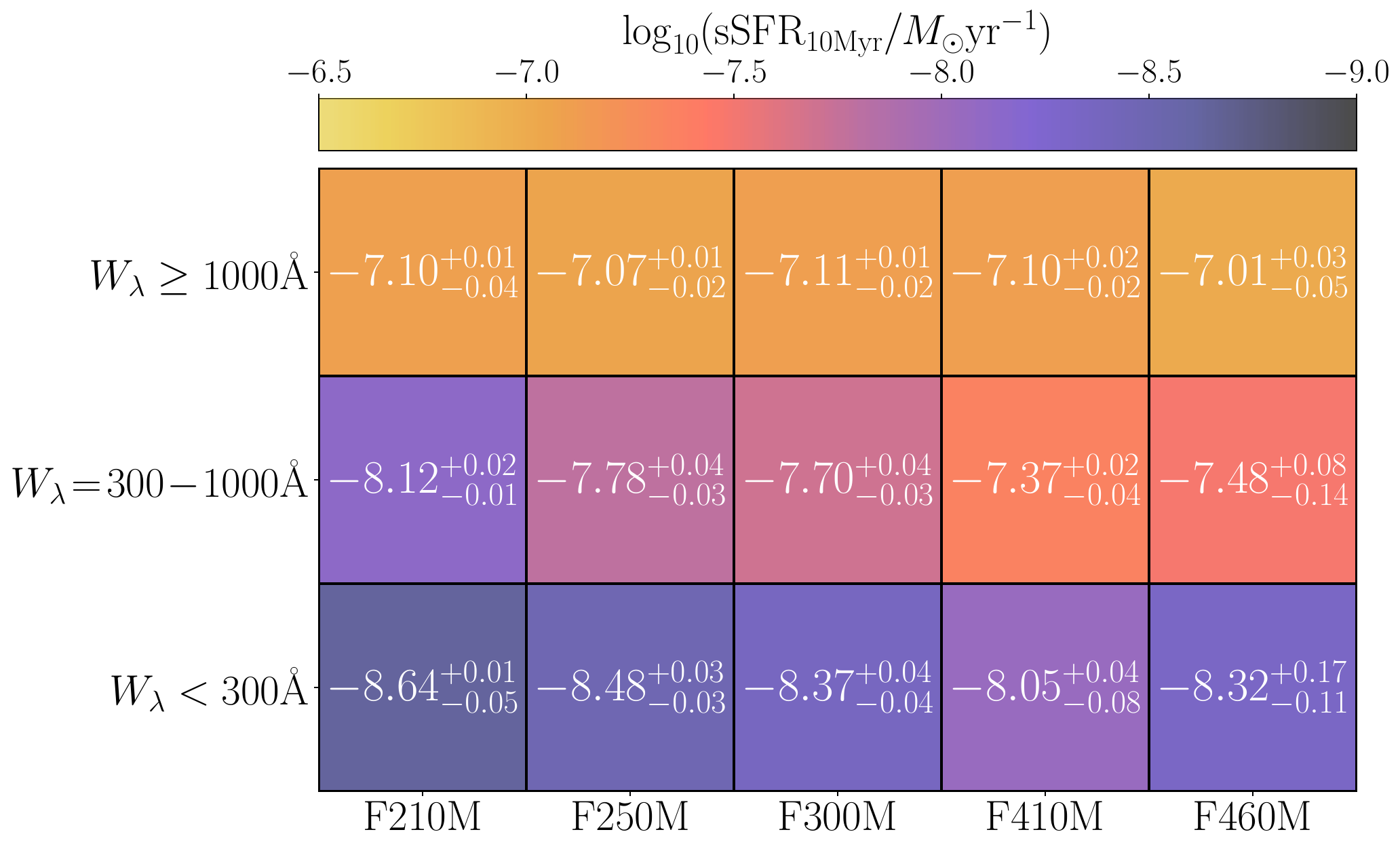}
    \caption{}
    \label{fig:ssfr}
    \end{subfigure}
    
    \begin{subfigure}[t]{0.48\textwidth}
    \centering
    \includegraphics[width = \textwidth]{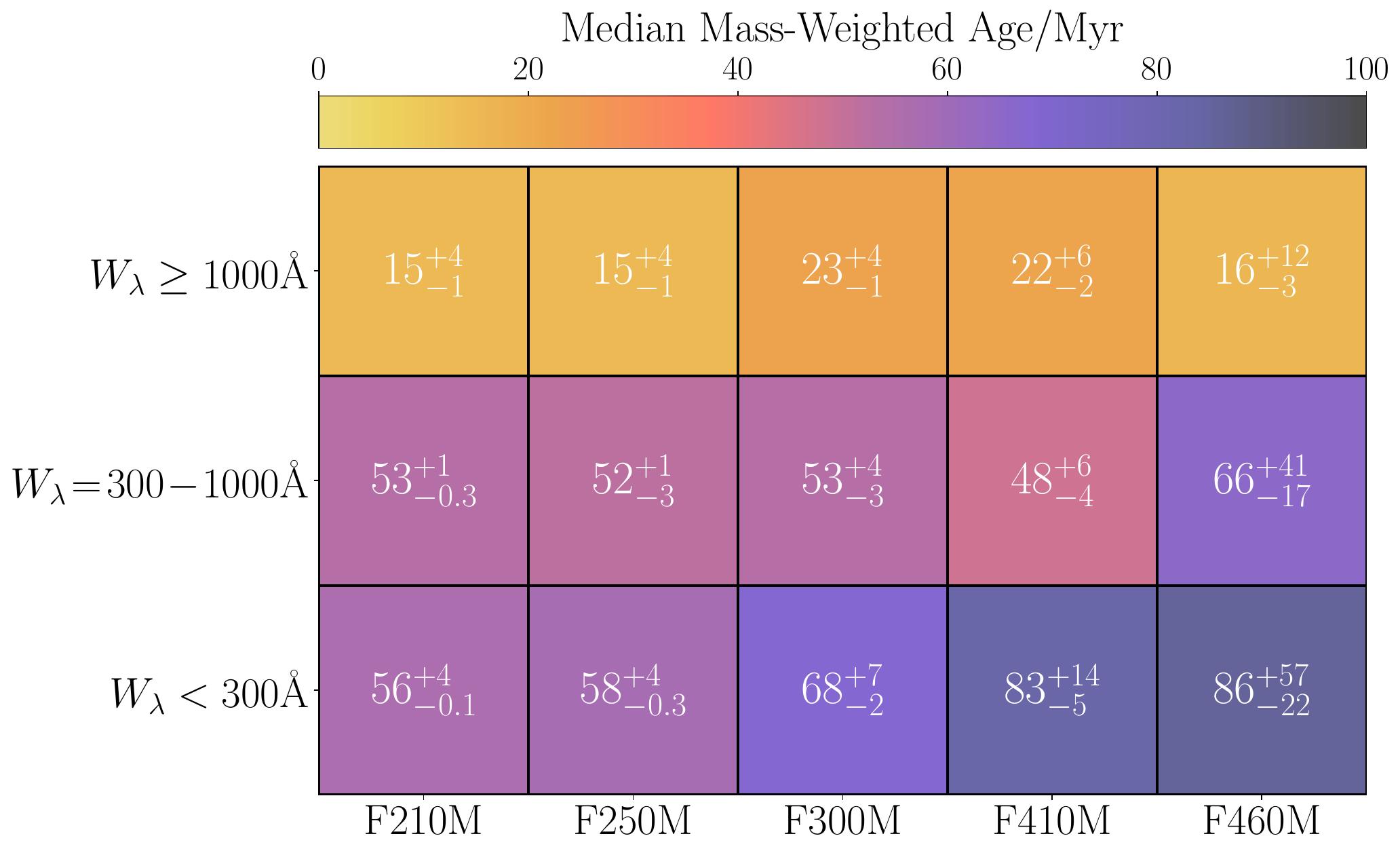}
    \caption{}
    \label{fig:mwa_indiv}
    \end{subfigure}
    \captionsetup{skip=0.001pt}
    \caption{The specific star-formation rate averaged over the last 10 Myr (top), and median mass-weighted age (bottom), for the F210M, F250M, F300M, F410M and F460M sub-samples (left-to-right), corresponding to $z\simeq3.2$, $z\simeq4.0$, $z\simeq 5.0$, $z\simeq7.2$ and $z\simeq8.3$. The samples are divided-up by equivalent width from top-to-bottom in each panel and the quoted uncertainties are the standard errors.}
    \label{fig:mwa}
\end{figure}
\subsection{Extreme $\mbox{[O\,\sc{iii}]} + \rm{H}\beta$ Emitters}
\label{sec:eelg}

The evolution in the fraction of so-called extreme emission-line galaxies, those galaxies displaying very high equivalent width rest-frame optical emission lines, has been investigated in a number of recent studies. Whilst rare in the local universe, \textit{JWST-}based studies have shown that the fraction of EELGs was much higher at earlier times \citep[e.g.][]{Sun_2023, Matthee_2023, 2024MNRAS.529.3301T, 2024MNRAS.535.1796B}, with their enhanced prevalence typically attributed to increasingly bursty star formation at high redshift.

Throughout our analysis, we define an EELG as an object with a rest-frame $\mbox{[O\,\sc{iii}]} + \rm{H}\beta$ equivalent width of $W_{\lambda}\geq1000\,$\AA. Across our five redshift sub-samples, for galaxies in the range $-21.6<M_{\mathrm{UV}}\leq -19.75$, this results in EELG fractions of $0.14^{+0.02}_{-0.02}$, $0.19^{+0.02}_{-0.02}$, $0.31^{+0.03}_{-0.03}$, $0.43^{+0.04}_{-0.04}$ and $0.50^{+0.10}_{-0.10}$. These EELG fractions are shown as filled black circles in Fig. \ref{fig:eelg}. In Fig. \ref{fig:eelg} we also show the EELG fractions for galaxies with $-21.6<M_{\mathrm{UV}}\leq -20.5$ as filled purple squares and galaxies with $-20.5<M_{\mathrm{UV}}\leq -19.75$ as filled orange squares. It can be seen that there is no indication that the evolution in the EELG fraction is a strong function of $M_{\rm UV}$. This is reassuring, because it implies that within the $M_{\rm UV}$ range we have chosen to study, our redshift sub-samples are not strongly biased towards including only high equivalent width sources.

For comparison with our results, in Fig. \ref{fig:eelg} we also show the EELG fractions measured by \citet{10.1093/mnras/staa3370}, \citet{10.1093/mnras/stac1109}, \citet{10.1093/mnras/stad1919}, \citet{10.1093/mnras/stae1857}, and \citet{2025MNRAS.537.3245B}, over a wide range of redshifts and at similar UV luminosities.
For the studies that adopted lower EELG $W_{\lambda}$ thresholds than employed here, or did not report EELG fractions (all except \citealt{2025MNRAS.537.3245B}), we modelled their $W_{\lambda}$ distributions as log-normal and calculated median, $\mu$, and standard deviation, $\sigma$, values. We then calculated the extreme-emitter fraction from this distribution, and repeated 1000 times. Taking the $50^{\rm th}$ and $16^{\rm th}/84^{\rm th}$ percentiles of the posterior distribution enabled us to obtain the median and $1\sigma$ errors on the EELG fractions. For the \citet{10.1093/mnras/stac1109} data, we applied an additional correction factor of $\times1.48$ to convert the $\mbox{[O\,\sc{iii}]}\lambda5007$ measurements into  $\mbox{[O\,\sc{iii}]} + \rm{H}\beta$ measurements \citep{2025MNRAS.537.3245B}. 

Combined with the existing literature results, our new set of self-consistent measurements add to the consensus that, despite the scatter between the results of different studies, the fraction of EELGs is a strongly increasing function of redshift. The robustness of this evolution, along with the various quadratic fits shown in Fig. \ref{fig:eelg}, are discussed in Section \ref{sec:EW_disc}.

\subsection{Stacked SEDs}

In order to explore differences in the galaxy SEDs as a function of equivalent width, in Fig. \ref{fig:stack_sed} we show stacked SEDs for our two highest redshift sub-samples ($6.7\leq z \leq 7.6$ and $8.0\leq z \leq 8.5$). To create the stacked SEDs, each object's \textsc{bagpipes} posterior SED distribution was sampled and the median flux at each wavelength taken to create a median spectrum. These spectra were then shifted to their rest-frame and normalised by the flux at $\lambda_{\rm{rest}} = 4000$\,\AA, after which the median was calculated to produce the final stacked SED. In Fig. \ref{fig:stack_sed}, stacked SEDs are presented for objects with $W_{\lambda}\geq 1000$\,\AA, $300 \AA \leq W_{\lambda}< 1000$\,\AA\, and $W_{\lambda}< 300$\,\AA, along with the corresponding stacked photometry.

The stacked EELG SEDs, shown in orange, correspond to very young stellar populations, as indicated by the strong Balmer jump. The Balmer jump seen in the stacked \textsc{bagpipes} SED is not obvious in the stacked photometry due to the presence of strong line emission immediately long-ward of the jump. The strong nebular emission, both continuum and line, becomes less prominent in the lower equivalent width SED stacks, with the $W_{\lambda}< 300$\,\AA\, stack instead showing a clear Balmer break. Overall, this is consistent with the lower $W_{\lambda}$ sources being systematically older and having lower specific star-formation rates.

This conclusion is supported by the results shown in Fig. \ref{fig:mwa}, where we present the \textsc{bagpipes}-derived median sSFR ($10\, \rm{Myr}$ average) and mass-weighted age of the five redshift sub-samples as a function of $\mbox{[O\,\sc{iii}]}+\mathrm{H}\beta$ equivalent width. Again, as one moves to lower $W_{\lambda}$ the galaxies become systematically older and display lower sSFRs. We will return to examine the robustness of this conclusion in Section \ref{sec:bursty}.

\subsection{$W_{\lambda}(\mbox{[O\,\sc{iii}]} + \rm{H}\beta)$ - Stellar Mass Correlation}

\begin{figure}
    \centering
    \includegraphics[width=\linewidth]{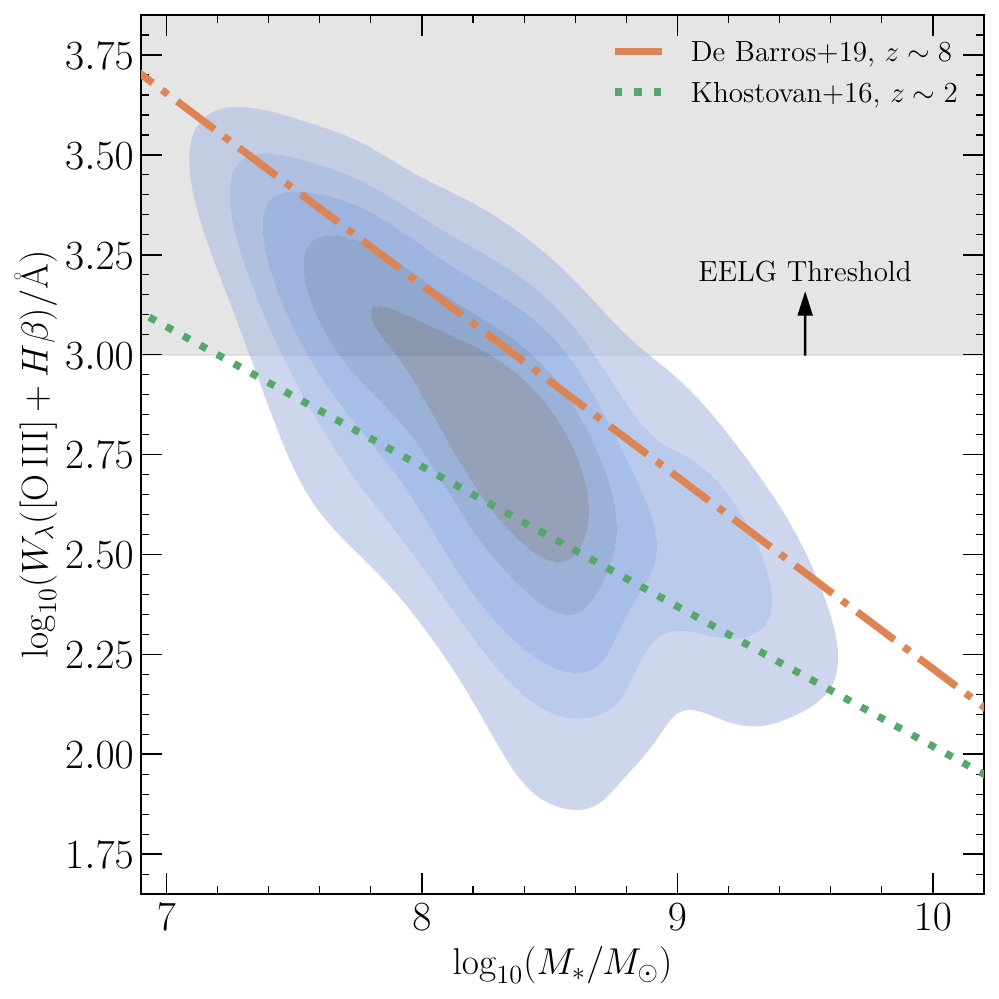}
    \caption{$W_{\lambda}(\mbox{[O\,\sc{iii}]} + \rm{H}\beta)$ versus stellar mass, with contours  showing a kernel density estimate visualisation of the overall data set. The power-law fits of \citet{10.1093/mnras/stz940} and \citet{2016MNRAS.463.2363K} are shown as dot-dash and dotted lines, respectively. We find that $\simeq 70\%$ of EELGs (grey shaded region) have stellar masses of $M_{\star}\lesssim10^{8}{\rm M}_{\odot}$.}
    \label{fig:properties}
\end{figure}

In Fig. \ref{fig:properties} we plot $\mbox{[O\,\sc{iii}]} + \rm{H}\beta$ equivalent width against stellar mass for our full $2.95\leq z \leq8.5$ sample, with the data displayed as contours to aid visualisation. 
Overall there is a relatively strong correlation between the two parameters, with a Pearson correlation coefficient of $r_{p} = -0.61$ ($p-$value of $<0.0001$). We do not fit a power-law to our sample because we do not expect it to be mass-complete, but our data set is consistent with, and bracketed by, the $z\simeq2$ and $z\simeq 8$ power-law fits of \citet{2016MNRAS.463.2363K} and \citet{10.1093/mnras/stz940}.
It is clear from Fig. \ref{fig:properties} that the EELG population (grey shaded area) is confined to low stellar masses, with $\simeq 70\%$ of EELGs having $\log_{10}(M_{\star}/{\rm M}_{\odot})<8.0$. 

\section{Discussion}
\label{sec:disc}

The results presented in Section \ref{sec:results} highlighted the redshift evolution of the $\mbox{[O\,\sc{iii}]} + \rm{H}\beta$ equivalent width and the accompanying increase in the fraction of EELGs, along with evidence that the physical driver for this evolution is the increasing burstiness of star formation in the young Universe. In this section, we provide a more detailed discussion of each of these key issues. 

\subsection{Redshift evolution of [\mbox{O\,\textsc{iii}}]$+\mathrm{H\beta}$ equivalent width}

\label{sec:EW_disc}

We show the redshift evolution of the [\mbox{O\,\textsc{iii}}]$+\mathrm{H\beta}$ equivalent width ($W_\lambda$) distribution from $z\simeq3$ to $z\simeq8$ in Fig. \ref{fig:ew_dist}. Across the five redshift sub-samples, we see a clear monotonic increase in the typical $W_\lambda$, rising from $\simeq350\,$\AA\, at $z\simeq 3.2$ to $\simeq1100\,$\AA\, at $z\simeq8.3$.

For our F410M sub-sample at $z\simeq 7.2$, we find a median equivalent width of $\mu=832^{+39}_{-37}\,$\AA\, ($\mathrm{log}_{10}(\mu)=2.92\pm0.02$), in excellent agreement with \citealt{10.1093/mnras/stad1919} ($\mu=780^{+70}_{-50}$\AA, $z\simeq7.3$). Our result is also consistent with, although marginally higher than, the $\mathrm{log}_{10}(\mu)=2.83^{+0.04}_{-0.03}$ obtained by \citet{2025MNRAS.537.3245B}. The small offset can be explained by the additional SED constraints afforded by the MINERVA medium bands, as well as the SFH model adopted here allowing for larger bursts (see Section \ref{sec:ew_methods}). We note that if we adopt the exact SFH prescription employed by \citet{2025MNRAS.537.3245B}, our $z\simeq7.2$ result changes to $\mathrm{log}_{10}(\mu)=2.78\pm0.02$.

The strong redshift evolution of [\mbox{O\,\textsc{iii}}]$+\mathrm{H\beta}$ equivalent width is also clearly visible in Fig. \ref{fig:eelg}, where we show the fraction of sources classed as extreme emission-line galaxies ($W_\lambda\gtrsim1000\,$\AA; $f_\mathrm{EELG}$) for the five redshift sub-samples. We observe a strong increase in $f_\mathrm{EELG}$, rising from $0.14\pm{0.02}$ at $z\simeq3.2$ to $0.50\pm{0.10}$ at $z\simeq8.3$. We find that the evolution in $f_\mathrm{EELG}$ is approximately quadratic in $(1+z)$, in agreement with past literature results \citep[e.g.][]{10.1093/mnras/staa3370,10.1093/mnras/stac1109,10.1093/mnras/stad1919,2025MNRAS.537.3245B}. An exception to this trend is seen in \citet{10.1093/mnras/stae1857}, where the $W_\lambda$ distribution median, and hence $f_\mathrm{EELG}$, actually decreases from $z\simeq5.6$ to $z\simeq7.3$.

\begin{table}
    \caption{Parameters of our quadratic fit ($f^{\rm{EELG}} = f_{0}^{\rm{EELG}}(1+z)^{2} + \gamma$) to the EELG fraction with redshift plotted in Fig. \ref{fig:eelg}. The parameters are listed for the two $M_{\rm UV}$ sub-samples and the combined sample.}
    \centering
    \renewcommand{\arraystretch}{1.3}
    \begin{tabular}{c|c|c}
    \hline
    $M_{\mathrm{UV}}$ Subsample & $f_{0}^{\rm{EELG}}$ & $\gamma$ \\
    \hline
    $-20.5<M_{\mathrm{UV}}\leq -19.75$ & $0.0063\pm0.0010$ & $0.04\pm0.03$ \\
    $-21.6<M_{\mathrm{UV}}\leq -20.5\phantom{0}$ & $0.0057\pm0.0014$ & $0.02\pm0.05$ \\
    \hline
    \makecell{\textbf{Combined}\\$-21.6<M_{\mathrm{UV}}\leq -19.75$} & $0.0062\pm0.0010$ & $0.04\pm0.03$ \\
    \hline
    \end{tabular}
    \label{tab:eelg}
\end{table}

In the selection of our redshift sub-samples, we implemented a SNR $\geq3$ threshold in the medium-band filter that sampled the [\mbox{O\,\textsc{iii}}]$+\mathrm{H\beta}$ complex. In principle, this could introduce an observational bias against the inclusion of the lowest $W_\lambda$ sources. However, it can be seen from Fig. \ref{fig:eelg} and Table \ref{tab:eelg} that, although there is a slight trend for the fainter $M_{\rm UV}$ sub-sample to display higher EELG fractions across our full redshift range, the difference is statistically marginal.

Due to the previous lack of deep, longer-wavelength, \textit{JWST} medium-band data, [\mbox{O\,\textsc{iii}}]$+\mathrm{H\beta}$ equivalent width estimates at $z\gtrsim8$ are scarce. Here, deep F460M imaging from MINERVA has enabled the most robust constraints to date of [\mbox{O\,\textsc{iii}}]$+\mathrm{H\beta}$ $W_\lambda$ at these redshifts, where we infer $\mathrm{log}_{10}(\mu)=3.05\pm0.03$ ($\simeq1120^{+80}_{-75}$\,\AA) from a sample of $N=49$ galaxies at $8.0\leq z \leq8.5$. This is in good agreement with \citet{2026arXiv260404284G}, who obtain $\mu \simeq 1260^{+330}_{-260}$\,\AA\, at $9.4\leq z <11.3$, using MIRI/F560W imaging, although notably from a fainter sample ($\langle M_\mathrm{UV}\rangle\simeq-19.3$) than our $z\simeq8.3$ sample ($\langle M_\mathrm{UV}\rangle\simeq-19.9$). We also note that the $z\simeq 8.3$ sample in this work is $16\times$ larger than the \citet{2026arXiv260404284G} sample, further highlighting the vital statistical robustness afforded by large-scale medium band programmes such as MINERVA.

\subsection{Evidence of Extremely Bursty Star-Formation}

\label{sec:bursty}

Our results provide multiple lines of evidence for increasingly bursty star-formation at $z\simeq5-9$. Firstly, the stacked SEDs shown in Fig.~\ref{fig:stack_sed} indicate that galaxies across the highest-redshift samples are dominated by young stellar populations, particularly amongst the most extreme emitters ($W_{\lambda}\geq1000$\,\AA). These sources exhibit both young mass-weighted ages and elevated sSFRs (Fig. \ref{fig:mwa}), consistent with intense episodes of recent star-formation following periods of reduced activity. Such rapidly varying SFHs are increasingly expected at early cosmic times, and are broadly consistent with both observational and simulation-based studies of high-redshift burstiness \citep[e.g.][]{Faisst_2019, Strait_2023,Sun_2023, 2024Natur.629...53L, looser2025jades}.

Further evidence is provided by the evolution in the shape of the $W_{\lambda}$ distributions shown in Fig. \ref{fig:ew_dist}. Galaxies observed during the peak of a star-formation burst are expected to exhibit strongly enhanced nebular emission and hence large $\mbox{[O\,\sc{iii}]}+\rm{H}\beta$ equivalent widths. Conversely, a bursty population should also contain galaxies observed during lower-SFR phases with substantially weaker emission lines \citep[see e.g.][]{2024Natur.629...53L, Faisst_2024, 10.1093/mnras/staf213}. In this work, at $z\simeq3.2$ the inferred $W_{\lambda}$ distribution is approximately symmetric, whereas progressively stronger asymmetry emerges towards higher redshift, becoming particularly pronounced by $z\simeq8.3$.

A similar departure from the approximately log-normal $W_{\lambda}$ distributions seen at lower redshift was identified by \citet{2025MNRAS.537.3245B}. While the typical equivalent width increases strongly with redshift, the increasingly extended tail towards low $W_{\lambda}$ may trace galaxies observed during temporary downturns in their recent star-formation activity. Such a distribution is qualitatively expected if elevated-SFR bursts are short-lived relative to the intervening lower-SFR phases, and therefore provides independent evidence for increasingly variable SFHs towards early cosmic times. Physically, the connection between $W_{\lambda}$ and burstiness is expected: nebular emission responds rapidly to changes in the instantaneous SFR, whereas the underlying stellar continuum traces star-formation over longer timescales. A rapid increase in SFR can therefore temporarily produce exceptionally large emission-line equivalent widths.

The stacked SED of the EELG subsample provides an additional signature of these young stellar populations, exhibiting a prominent Balmer jump at $\lambda_{\rm rest}\simeq3645$\,\AA\, (Fig. \ref{fig:stack_sed}). This feature, which is mainly detected through SED modelling, arises from nebular free-bound emission and becomes particularly strong in young, nebular-dominated systems \citep{10.1093/mnras/stae1547, Katz202521}. A weaker ($\lesssim0.5\times$) discontinuity is also evident in the 300\,\AA\ $\leq W_{\lambda}<1000$\,\AA\, subsample, consistent with somewhat older but still young stellar populations. In both cases, however, the Balmer jump is less apparent in the stacked photometry because of the strong contribution from emission lines red-ward of the discontinuity.

Although $\mbox{[O\,\sc{iii}]}+\rm{H}\beta$ equivalent width provides a sensitive tracer of recent changes in star-formation activity, low $W_{\lambda}$ does not uniquely imply a low instantaneous SFR. Recent studies have identified galaxies for which standard low-density direct-method assumptions imply extremely low metallicities, but where substantially higher electron densities alter the inferred gas conditions and metallicities \citep[][it should be noted that these studies focus on small numbers of extreme objects]{2025ApJ...982...14H, 2026arXiv260820339H, Berg_2026, 2026MNRAS.547ag380A}. At sufficiently high densities, collisional de-excitation can suppress $\mbox{[O\,\sc{iii}]}$ emission, producing weaker observed $\mbox{[O\,\sc{iii}]}+\rm{H}\beta$ equivalent widths than would otherwise be expected. Some fraction of the low-$W_{\lambda}$ population could therefore comprise actively star-forming, higher-metallicity systems in which the $\mbox{[O\,\sc{iii}]}$ emission is suppressed, rather than galaxies observed exclusively during post-burst phases or extremely metal-poor galaxies. 
Despite the data quality available to this study, it is clear that resolving this issue requires deep, medium resolution ($R\simeq 1000$), NIRSpec spectroscopy.

\subsection{Associated Physical Properties}

The results presented in Fig. \ref{fig:properties} show a clear anti-correlation between $\mbox{[O\,\sc{iii}]}+\rm{H}\beta$ equivalent width and stellar mass. Although we do not fit a power-law relation to our sample due to mass-completeness concerns, our results are 
consistent with previous observational studies \citep[e.g.][]{Reddy_2018, 10.1093/mnras/stz2236, 2022MNRAS.511.4464A, 2025MNRAS.537.3245B} and the shape of the contours is in line with the $(M_\ast/{\rm M_\odot)}^{-0.48}$ scaling (dot-dash line) inferred by \citet{10.1093/mnras/stz940}. 

A correlation between $\mbox{[O\,\sc{iii}]}+\rm{H}\beta$ equivalent width and stellar mass trend is expected if lower-mass galaxies experience increasingly variable star-formation histories, with burstiness becoming more prominent towards both lower stellar masses and higher redshifts \citep[e.g.][]{looser2025jades}. In fact, $\simeq70\%$ of our EELG sample have stellar masses of $M_{\star}\lesssim10^{8}{\rm M}_{\odot}$ and $\simeq90\%$ have stellar masses of $M_{\star}\lesssim10^{8.5}{\rm M}_{\odot}$.

At fixed $M_\star$, however, we recover substantial scatter in $W_\lambda$, with several effects likely contributing to this dispersion. For example, in the absence of longer-wavelength constraints from MIRI, the stellar masses in the two highest-redshift bins remain relatively uncertain due to the potential impact of outshining \citep[e.g.][]{conroy2010}.
Some fraction of the scatter may also be intrinsic, with galaxies of similar stellar mass exhibiting substantially different equivalent widths, depending on whether they are observed during an elevated or suppressed phase of their recent SFH. Finally, covariance between the inferred stellar mass and emission-line strength within the SED fitting can further broaden the observed relation.

The combined PRIMER+MINERVA dataset also enables us to investigate the ionizing photon production efficiency, $\xi_{\rm ion}$, across $z\simeq3-8$. We infer $\xi_{\rm ion}$ using the empirical relation from \citet{begley26}, together with the conversion between $\mbox{[O\sc{iii}]}+\rm{H}\beta$ and $\mbox{[O\sc{iii}]}\lambda5007$ equivalent widths described in Section \ref{sec:eelg}.

\begin{figure}
    \centering
    \includegraphics[width=\linewidth]{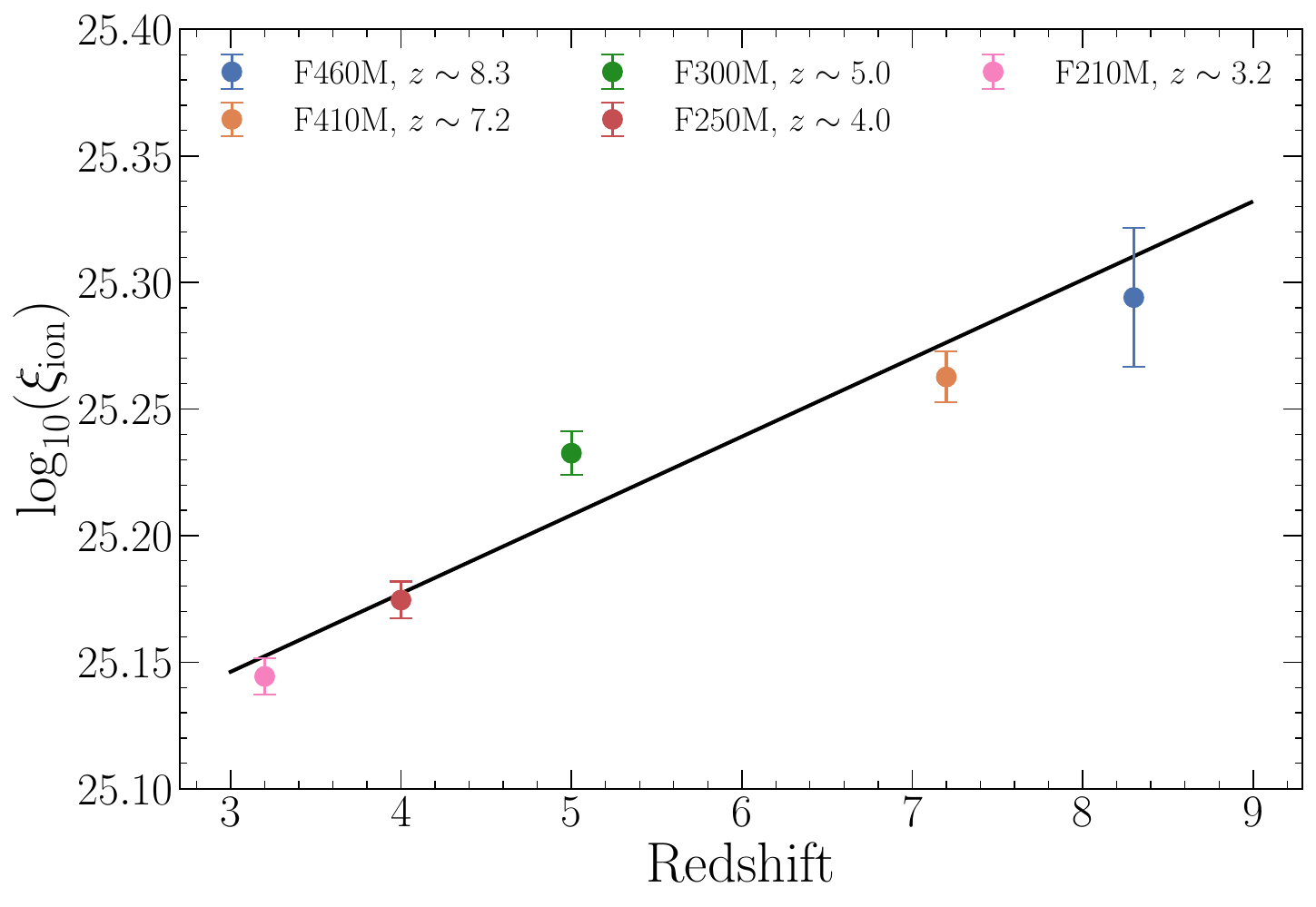}
    \caption{The correlation between $\rm{log}_{10}(\xi_{ion})$ and redshift, showing the median ionising photon production efficiency for each sub-sample. The best-fitting relation (black line) has the form: $\mathrm{log}_{10}(\xi_{\mathrm{ion}}) = (0.031\pm0.003)z + (25.053\pm0.013)$.}
    \label{fig:xi_ion}
\end{figure}

We recover median $\mathrm{log}_{10}(\xi_{\rm ion} \,/\mathrm{Hz^{-1}\,erg})$ values of $25.14\pm0.01$, $25.17\pm0.01$, $25.23\pm0.01$, $25.26\pm0.01$ and $25.29\pm0.03$ across the F210M, F250M, F300M, F410M, and F460M samples, respectively (Fig. \ref{fig:xi_ion}). These measurements imply only mild evolution with redshift, with $d\,\mathrm{log}_{10}(\xi_{\rm ion})/d\,\mathrm{log}_{10}(z)=0.031\pm0.003$. This is shallower than the evolution reported by \citet{2024MNRAS.527.6139S} and \citet{2022ApJ...935...94S}, but broadly consistent with more recent measurements \citep[e.g.][]{2024MNRAS.535.2998S, 2025MNRAS.537.3245B}. Indeed, \citet{2025MNRAS.537.3245B} showed that increasingly bursty SFHs towards high redshift can naturally flatten the apparent redshift evolution of $\xi_{\rm ion}$. We emphasise, however, that $\xi_{\rm ion}$ depends on multiple correlated galaxy properties, and the weak evolution recovered here is therefore unlikely to reflect redshift alone \citep{begley26}.

\section{Conclusions}
\label{sec:conclusions}

In this work, we explored the scientific gains enabled by combining \textit{JWST}/NIRCam imaging from the PRIMER and MINERVA surveys in the UDS field. Together with the available \textit{HST}/ACS imaging, the combined catalogue provides photometry across 19 broad- and medium-band filters, significantly improving the constraints provided by SED fitting. Using \textsc{bagpipes}, with non-parametric SFHs, we modelled a sample of $N=1964$ galaxies with $M_{\rm UV}\leq-19.25$ at $z\simeq3-8.5$ and inferred their $\mbox{[O\,\sc{iii}]}+\rm{H}\beta$ equivalent widths. 
Based on these measurements, we investigated the redshift evolution of $\mbox{[O\,\sc{iii}]}+\rm{H}\beta$ equivalent width and the relationship between equivalent width, age, stellar mass, and star-formation rate. Our main conclusions are as follows:

\begin{enumerate}
    \item Based on taking the median of 5 photometric redshift runs, we find that the combined PRIMER+MINERVA data set is capable of delivering highly accurate photometric redshift estimates. When compared to a robust sample of $N=3145$ spectroscopic redshifts, drawn from ground-based surveys and the DJA, our final photometric redshifts have an outlier rate of $\eta = 4.71\%$ and $\sigma=0.015$, over the full redshift range and down to a magnitude limit of $m_{\rm F444W}\leq28$.
    This represents a $42\%$ reduction in outliers compared to photometric redshifts based on PRIMER data alone. When combined with a cut based on the standard deviation of the 5 photometric redshift runs, we showed that it is possible to reduce the outlier rate to $\eta = 2.69\%$, while culling $<7\%$ of the sample.    

    \item After exploring a variety of different star-formation history prescriptions with \textsc{bagpipes}, we determined that fitting the PRIMER+MINERVA photometry with BPASS SPS models and an adaptation of the \citet{Leja_2019} non-parametric continuity SFH produced the most accurate recovery of the $\mbox{[O\,\sc{iii}]} + \rm{H}\beta$ equivalent width. 

    \item As a final consistency check on our sample selection and equivalent width methodology, we presented new constraints on the bright-end of the
    $\mbox{[O\,\sc{iii}]} + \rm{H}\beta$ luminosity function at $z\simeq7.3$. Our results are in good agreement with \citet{2025ApJ...980..200W, 2026arXiv260605323W} and are consistent with those of \citet{10.1093/mnras/stae2353} and \citet{2026A&A...712A..16M}. By combining our bright-end data with the fainter data of \citet{2026arXiv260605323W}, we fit a Schechter function with the following parameters: $\mathrm{log}_{10}(\phi^{*}) = -4.3_{-0.5}^{+0.2}\, \rm{Mpc^{-3}\,dex^{-1}}$, $\mathrm{log}_{10}(L^{*}) = 43.4_{-0.2}^{+0.3}\, \rm{erg\, s^{-1}\,dex^{-1}}$, and $\alpha = -2.15^{+0.05}_{-0.07}$. 
    
\item Based on a consistent analysis of five redshift sub-samples spanning the redshift range $z=3-8.5$, we find that the median $\mbox{[O\,\sc{iii}]}+\rm{H}\beta$ equivalent width increases monotonically with redshift, rising from $\simeq350\,$\AA\, at $z\simeq 3.2$ to $\simeq1100\,$\AA\, at $z\simeq8.3$. These results are broadly consistent with, although more comprehensive than, previous measurements \citep[e.g.][]{2025MNRAS.537.3245B, 10.1093/mnras/stz940, 10.1093/mnras/stad1919}.

\item In addition to the evolution in median equivalent width, we find that the equivalent width distribution becomes increasingly asymmetric with redshift. As a result, we see a roughly $(1+z)^2$ increase in the fraction of EELGs, rising from $0.14\pm{0.02}$ at $z\simeq3.2$ to $0.50\pm{0.10}$ at $z\simeq8.3$. This increase in the EELG fraction agrees well with previous constraints \citep[e.g.][]{10.1093/mnras/staa3370, 10.1093/mnras/stad1919, 2025MNRAS.537.3245B, 2026arXiv260404284G}.

\item Stacked SEDs of the F460M and F410M samples were used to explore the differences between extreme emission-line galaxies and weaker emitters. This confirmed that EELGs are consistent with having very young stellar populations, as indicated by the presence of a strong Balmer jump. In contrast, those galaxies with low equivalent widths (i.e. $W_{\lambda}<300\,$\AA), are consistent with older stellar populations, as indicated by the presence of a well developed Balmer break.

\item The conclusion that the prevalence of EELGs at high redshift is connected with increasingly bursty star formation is further supported by considering the median mass-weighted age and specific star-formation rate of our galaxy sample over the 10 Myr prior to observation. This confirms a consistent link between higher $\mbox{[O\,\sc{iii}]}+\rm{H}\beta$ equivalent width, high specific star-formation rate and young mass-weighted age. Over the full redshift range of our study, we find that EELGs are confined to a mass-weighted age of $<23$ Myr and $\log_{10}({\rm sSFR}/{\rm Gyr})\geq -7.1$.

\end{enumerate}

\section*{Acknowledgements}

The authors would like to thank Johann Cohen-Tanugi and Olivier Ilbert for their help in improving the photometric redshift fitting through \textsc{LePhare}. IJBH acknowledges the support of the Science and Technology Facilities Council. RJM and TMS acknowledge support from a Science and Technology Facilities Council Grant (grant reference ST/Y001095/1). DJM and JSD acknowledge the support of the Royal Society through the award of a Royal Society Research Professorship to JSD. ACC, HL, ET, and SDS acknowledge support from a UKRI Frontier Research Grantee Grant (PI Carnall; grant reference EP/Y037065/1). FC acknowledges support from a UKRI Frontier Research Guarantee Grant (PI Cullen; grant reference EP/X021025/1).
This paper is based on observations made with the
NASA/ESA/CSA James Webb Space Telescope. The data were
obtained from the Mikulski Archive for Space Telescopes at the
Space Telescope Science Institute, which is operated by the Association of Universities for Research in Astronomy, Inc., under
NASA contract NAS 5-03127 for JWST. Some
of the data products presented herein were retrieved from the Dawn
JWST Archive (DJA). DJA is an initiative of the Cosmic Dawn Center, which is funded by the Danish National Research Foundation
under grant No. 140. This work made use of v2.2.1 of the Binary Population and Spectral Synthesis (BPASS) models as described in \citet{Eldridge_Stanway_Xiao_McClelland_Taylor_Ng_Greis_Bray_2017} and \citet{10.1093/mnras/sty1353}. This research has made use of the SVO Filter Profile Service
(http://svo2.cab.inta-csic.es/theory/fps/) supported from the Spanish
MINECO through grant AYA2017-84089 \citep{2012ivoa.rept.1015R, 2020sea..confE.182R}. This research has made use of NASA’s Astrophysics Data System Bibliographic Services.

\section*{Data Availability}

At the time of publication, all data relating to the images used in this work are available from the public Barbara A. Mikulski Archive for Space Telescopes (MAST).

\bibliographystyle{mnras}
\bibliography{oiii.bib}

\bsp	% typesetting comment
\label{lastpage}
\end{document}